\documentclass[10pt, a4paper]{article}
\usepackage[final]{lrec-coling2024} 
\usepackage{tipa}

\title{Sonos Voice Control Bias Assessment Dataset:\\A Methodology for Demographic Bias Assessment\\ in Voice Assistants}

\name{Chloé Sekkat\textsuperscript{1}, Fanny Leroy\textsuperscript{*}\thanks{\textsuperscript{*}Independent methodologist in statistics}, Salima Mdhaffar\textsuperscript{2}, Blake Perry Smith\textsuperscript{1},\\ {\bf \large Yannick Estève\textsuperscript{2}, Joseph Dureau\textsuperscript{1}, Alice Coucke\textsuperscript{1}}} 

\address{\textsuperscript{1}Sonos Inc., France \\\textsuperscript{2}Laboratoire Informatique d’Avignon, Avignon Université, France \\
         \textsuperscript{1}\{first.last\}@sonos.com \\
         \textsuperscript{2}\{first.last\}@univ-avignon.fr\\}

\abstract{
Recent works demonstrate that voice assistants do not perform equally well for everyone, but research on demographic robustness of speech technologies is still scarce. This is mainly due to the rarity of large datasets with controlled demographic tags. This paper introduces the Sonos Voice Control Bias Assessment Dataset, an open dataset composed of voice assistant requests for North American English in the music domain ($1,038$ speakers, $166$ hours, $170$k audio samples, with $9,040$ unique labelled transcripts) with a controlled demographic diversity (gender, age, dialectal region and ethnicity). We also release a statistical demographic bias assessment methodology, at the univariate and multivariate levels, tailored to this specific use case and leveraging spoken language understanding metrics rather than transcription accuracy, which we believe is a better proxy for user experience. To demonstrate the capabilities of this dataset and statistical method to detect demographic bias, we consider a pair of state-of-the-art Automatic Speech Recognition and Spoken Language Understanding models. Results show statistically significant differences in performance across age, dialectal region and ethnicity. Multivariate tests are crucial to shed light on mixed effects between dialectal region, gender and age.
 \\ \newline \Keywords{speech corpus, automatic speech recognition, spoken language understanding, demographic bias, bias detection} 
}

\begin{document}

\maketitleabstract

\section{Introduction}

One of the main applications of Automatic Speech Recognition (ASR) is Spoken Language Understanding (SLU). SLU combines speech and natural language processing techniques and is used in voice interfaces such as domestic voice assistants in smart speakers. It typically combines several tasks like intent classification and slot filling~\citet{tur2011slu}. Such a field of application raises many challenges, especially around footprint and privacy (always-on setting, edge processing like in~\citet{he2019streaming, saade2019spoken}), proper nouns and rare words recognition (e.g. for music entities recognition, in~\citet{li2021improving, sainath2021efficient}) and acoustic robustness (far-field noisy settings typical of the home environment, like in~\citet{Braithwaite2019SpeechEW, Dfossez2020RealTS}).

However less research is directed on what we could call \textit{demographic robustness} of voice assistants. There is some evidence though that ASR systems do not perform equally well for everyone (see Section~\ref{sec:related-work}). In particular, performance degradation stemming from demographic factors -- gender, age, accent, race -- can be observed, leading to our definition of demographic bias: when the performance of a speech recognition system depends on the group of people it is evaluated upon. 
Such research is nonetheless mostly focused on conversational use cases of the general domain for dictation and the underlying metric to optimize is transcription accuracy or Word Error Rate (WER). We believe these are not representative of voice assistant usage that focuses on the recognition of short action-oriented commands and that SLU metrics are likely to be a better proxy for voice assistant user experience. The correct execution of voice assistant requests indeed strongly depends on the recognition of isolated and sometimes complex entities, rather than on the exact transcription of the entire utterance.

In this work, we propose an open dataset, the Sonos Voice Control Bias Assessment Dataset ($1,038$ speakers, $166$ hours, $170$k audio samples, with $9,040$ unique labelled transcripts), composed of voice assistant requests in the music domain with a controlled demographic diversity (gender, age, dialectal region and race). The dataset is accompanied by a statistical bias assessment methodology tailored to this use case, leveraging intent and entity tagging through the Exact Match (EM) metric rather than WER. The statistical tests include a standard univariate approach, but also multivariate models to identify possible mixed effects. The proposed approach is illustrated on state-of-the-art End-to-End (E2E) ASR and SLU models. The code to reproduce our results is available on GitHub\footnote{\url{https://github.com/sonos/svc-demographic-bias-assessment}}.

\section{Related Work}
\label{sec:related-work}

\subsection{Bias Assessment}

\textbf{Existence of biases.} 
A number of works investigate the disparities in performance of commercial and open source ASR systems across a variety of speaker characteristics, most of them focusing on variations of English. 
Among the most studied demographic variables are native vs. non-native accents (dialectal region), gender, age, and race.

Evidence is mixed regarding gender: some studies found that ASR systems favor female speakers~\citet{koenecke2020racial, goldwater2010words, adda2005speech, sawalha2013effects, feng2023towards} while others found that male speakers have a better recognition rate~\citet{tatman2017gender, garnerin2019gender, garnerin2021investigating}, or no significant bias at all~\citelanguageresource{meyer2020artie}. 

ASR systems tend to better recognize native speakers vs. non-native ones~\citet{koenecke2020racial, palanica2019you, wu2020see,tatman2017effects, tatman2020sociolinguistic, feng2021quantifying, feng2023towards} due chiefly to language variability (regional and/or socio-linguistic), accents, articulation and speech rates. Evidence of commercial ASR systems exhibiting racial biases can be found in~\citet{koenecke2020racial, tatman2017gender} between Black and White speakers. Finally, several studies point out that younger speakers ($18-30$) are better understood than children and seniors~\citet{sawalha2013effects, feng2021quantifying}. This can be explained by the challenge represented by child speech due to their shorter vocal tracts, more variable speaking rate and inaccurate articulation~\citet{qian2017bidirectional}. 


\textbf{Bias assessment methodology.} Most of previous work focus on the average relative degradation of the transcription accuracy (through word, character or phoneme error rates) across speaker groups (e.g. male vs. female, native vs. non-native, White vs. African-American). To the best of our knowledge, no previous work aims at quantifying demographic bias through SLU metrics.

Only a handful of prior studies propose univariate statistical tests or models to find out whether these mean variations are statistically significant. Statistical tests such as Wilcoxon Rank Sum, Kruskall-Wallis~\citet{garnerin2021investigating, garnerin2019gender} and one-way analyses of variance (ANOVA)~\citet{feng2023towards, meyer2020artie} are used. Even fewer studies consider second order effects with, for instance, linear mixed-effects regressions~\citet{tatman2017gender, dichristofano2022performance} (with speaker and year as random effects) and mixed-effects Poisson regressions~\citet{liu2022model} (with speaker as random effect and demographic tag of interest as fixed effect). 

\subsection{Available datasets}

There is a lack of standard benchmarks in the literature for demographic bias assessment in voice assistants~\citet{ngueajio2022hey}. Though previous studies are extremely valuable, the datasets used are often small in terms of number of audio samples, speakers and transcripts variability. For instance, the one used in~\citet{tatman2017gender} is made of 62 words read in isolation by 80 speakers. Other studies use internal datasets such as \textit{VoiceCommand} in~\citet{liu2022model}. It was collected through a crowd-sourcing campaign  where a limited number of 95 participants were instructed to utter voice commands. When not limited by the number of speakers, the datasets are not representative of the voice assistant use-case but instead comprise books or stories~\citetlanguageresource{panayotov2015librispeech, bradlow2010allsstar}, broadcast news~\citelanguageresource{kalluri2021nisp}, conversational speech~\citetlanguageresource{oostdijk2000spoken, coraal, pitt2007buckeye}, human-machine interaction speech~\citetlanguageresource{cucchiarini2006jasmin}, interviews~\citetlanguageresource{hazirbas2021towards}, random paragraphs~\citetlanguageresource{weinberger2011speech}, and improvised speech~\citetlanguageresource{wang2021voxpopuli, bradlow2010allsstar}. 

One of the most recent attempts to publish an open source (under the Creative Commons CC-0 license) dataset for demographic bias assessment is the Artie Bias Corpus introduced in~\citetlanguageresource{meyer2020artie}, a manually annotated subset of the Common Voice~\citelanguageresource{Ardila2019CommonVA} test set comprising $1,712$ audio clips ($\approx 2.4$ hours), $1,903$ utterances, 3 gender classes, 8 age ranges and 17 English accent classes. However, the metadata is self reported and sometimes incomplete (the accent label is missing for around $33\%$ of the data).
This dataset is also very imbalanced towards younger, male, US English speakers. As pointed out by the authors, one of the main limitations of this dataset is its small size which has a direct impact on the statistical power of the tests.

\section{Sonos Voice Control Bias Assessment Dataset}
\label{sec:dataset}

The Sonos Voice Control Bias Assessment dataset addresses the usage of a voice assistant for music control, which is reported as the most common use case\footnote{\href{https://blog.adobe.com/en/publish/2018/09/06/adobe-2018-consumer-voice-survey}{Adobe Digital insights 2018}}, and includes $170,413$ audio samples in North American English ($\approx$~$166$ hours) along with their transcripts, labels, and demographic metadata about the speaker. The dataset was designed mostly for evaluation purposes, but splits for training and development are also made available. Audio samples were obtained following a directed process (read speech) described in the following sections, based on transcript and label production on the one hand, and on speaker specification and selection on the other hand. The dataset is available for download and can be used for academic and/or research purposes\footnote{\url{https://github.com/sonos/svc-demographic-bias-assessment}}.

\subsection{Transcripts and labels}
\label{sec:transcripts}

Transcripts of this dataset are divided into a set of 32 \textit{intents}, each standing for a class of actions which can typically be requested within the music domain, related to playing music, controlling the music stream, requesting information, switching, grouping and ungrouping devices, and managing the user's library. Intents can further specify optional or required \textit{entities}, in the form of slots, to make their semantic interpretation, in terms of action, complete and coherent. In the case of \texttt{PlayMusic} or \texttt{VolumeUp}, such entities can be names and titles from the music domain, or level quantifiers, identified as e.g. \texttt{artist} or \texttt{volume\_level}, which fill up the action of content playing with an actual, quantifiable or identifiable value. The music entities (\texttt{artist}, \texttt{song}, \texttt{album}) contain much more distinct values than the rest of the entities, with respectively $1,715$ artists, $2,643$ songs and $1,626$ albums values found in all the transcripts. There are 32 unique entities in the dataset, distributed among all intents. Transcripts and their labels (intents and slots) are generated through a semi-automated process that is beyond the scope of the present paper.

Two assistant interaction contexts are usually considered. On one hand, the \texttt{PlayMusic} intent (e.g. \textit{play the song Abbey Road}) which is the only one related to content initiation, with music entities and arguably the most complex intent: a third of the \texttt{PlayMusic} queries have 3 slots or more. On the other hand, the rest of the 31 intents are denoted \texttt{Transport Control} intents, e.g. \texttt{Forward} (\textit{fast forward a little}), \texttt{Again} (\textit{once more}), or \texttt{VolumeUp} (\textit{play louder}). They are typically short queries, $80\%$ of them have either 1 or no slot, and the number of possible phrasings is limited. There is a total of $9,040$ unique transcripts in the dataset, $8,114$ for the \texttt{PlayMusic} intent and $926$ for \texttt{Transport Control} intents. Appendix~\ref{subsec:appendix-ontology} provides provides more insights on the ontology coverage.

\subsection{Speaker demographic metadata}
A total of $1,038$ speakers were selected following three dimensions of demographic characteristics: \textbf{gender} (male and female here, though we recognize that gender cannot be captured by a binary variable), \textbf{age range} (5 age ranges starting at 9 years old) and \textbf{dialectal region}. The latter dimension accounts for dialectal variation, e.g. pronunciation and intonation contrasts and fluctuations inside the United States.  Six regional groups represent native speakers of American English and two groups represent non-native speakers residing in the USA, native of Asia (Asian) and Latin America (LatinX), with internal variations being expected within each region or group. More details on the choice and definition of these groups can be found in Appendix~\ref{subsec:appendix-dialectal-regions}.

\textbf{Ethnicity} was unreported in the initial version of the dataset. Though we made sure that regions (and more specifically cities) where ethnic diversity is prominent were represented in the data collection, this lack of information made it impossible to quantify bias between different racial groups. Some context can be found in Appendix~\ref{subsec:appendix-ethnicity}. We therefore launched an additional campaign targeting specifically Caucasians ($50$ speakers) and African Americans ($48$ speakers), present only in the test split. The gender distribution for these speakers is balanced ($49\%-51\%$) and the regional distribution is roughly similar to that of the rest of the dataset. However, recruiting children under 16 proved difficult, therefore constraints imposed on that age group were relaxed.

\subsection{Audio samples}
 
Each of the $1,038$ speakers have recorded $193$ distinct transcripts on average, except for 9 to 16 year old children ($328$ speakers) who recorded only $96$ transcripts on average and for speakers with ethnicity information ($98$ speakers) who recorded $150$ transcripts on average. This disparity is explained by the difficulties encountered when hiring speakers from these specific demographic groups. Recording conditions are uniformly clean and close-field. 

The total number of audio samples is $170,413$ over $9,040$ unique transcripts, ($\approx$~$166$ hours): $77,515$ ($92.88$ hours) for the \texttt{PlayMusic} intent and $92,898$ ($72.66$ hours) for the \texttt{Transport Control} intents. Train, dev and test splits are finally created by splitting speakers among $10$ speaker groups, each balanced in terms of demographic characteristics. Speaker groups 1 to 5 are used for the train ($428$ speakers, $69,206$ samples) and dev ($38$ speakers, $6,703$ samples) splits and speaker groups 6 to 10 for the test split ($572$ speakers, $94,504$ samples). A description of the splits can be found in Table~\ref{tab:stats-data}.

\begin{figure}[!ht]
\begin{center}
\includegraphics[scale=0.2]{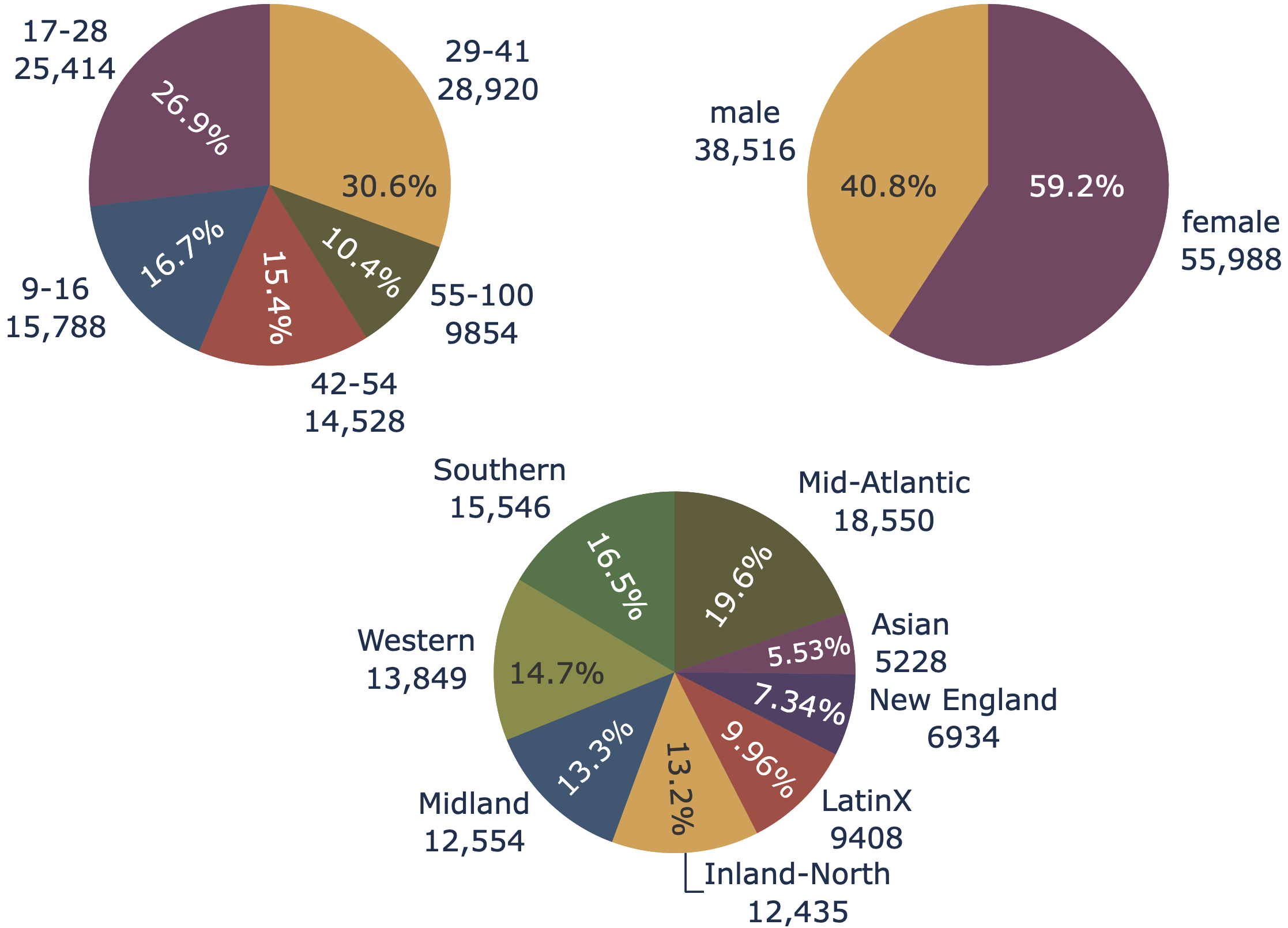} 
\caption{Audio sample distribution in the test split of the dataset in terms of age, gender, and dialectal region. The number of samples in each group is displayed under the group label.}
\label{fig:samples-distrib-test}
\end{center}
\end{figure}

Fig.~\ref{fig:samples-distrib-test} displays the audio sample distribution in the test split of the dataset for each demographic group. We see that the dataset is skewed towards female and younger speakers. We also note that the LatinX and Asian groups are slightly less populated than the other dialectal regions.

\begin{table}
\centering
\begin{tabular}{cccc}
\toprule
\textbf{Split} & \textbf{Samples} & \textbf{Speakers} & \textbf{Duration}\\
\midrule
Train & $69,206$ & $428$ & $68$:$49$\\
Dev & $6,703$ & $38$ & $6$:$26$\\
Test & $94,504$ & $572$ & $90$:$30$ \\
\bottomrule
\end{tabular}
\caption{\label{tab:stats-data} Description of the dataset across splits in terms of number of samples, number of speakers and duration (h:min).}
\end{table}

The 3 demographic dimensions included in this dataset (gender, age, dialectal region) amount to 80 different demographic groups (not considering the smaller subset with ethnicity labels), that we tried to keep reasonably balanced by a minimum speaker distribution in each of them. As an illustration, Table~\ref{tab:confusion-bias} displays the age and gender distribution for each dialectal region in the test split. We believe that the coverage in terms of speakers and audio samples in the Sonos Voice Control dataset should allow to assess potential biases of a voice assistant related to demographic characteristics of speakers. Further descriptive statistics can be found in Appendix~\ref{sec:appendix-svc-dataset-stats-desc} and ~\ref{sec:appendix-confusion-bias}.

\begin{table}[h!]
\centering
\resizebox{\columnwidth}{!}{%
\begin{tabular}{ccccc}
\toprule
\textbf{Dialectal region} & \textbf{Samples} & \textbf{Speakers} & \textbf{Age distribution} & \textbf{Gender distribution} \\ \midrule

Southern & $15546$ & $97$ & \begin{tabular}[c]{@{}c@{}}$9-16: 17\%$\\ $17-28: 23\%$\\ $29-41: 34\%$\\ $42-54: 15\%$\\ $55-100: 11\%$\end{tabular} &
  \begin{tabular}[c]{@{}c@{}}Female: $55\%$\\ Male: $45\%$\end{tabular} \\ \midrule
  
Western & $13849$ & $87$ & \begin{tabular}[c]{@{}c@{}}$9-16: 19\%$\\ $17-28: 28\%$\\ $29-41: 24\%$\\ $42-54: 20\%$\\ $\textbf{55-100: 9\%}$\end{tabular} & \begin{tabular}[c]{@{}c@{}}Female: $61\%$\\ Male: $39\%$\end{tabular} \\ \midrule

Inland North & $12435$ & $73$ & \begin{tabular}[c]{@{}c@{}}$9-16: 10\%$\\ $17-28: 24\%$\\ $29-41: 34\%$\\ $42-54: 18\%$\\ $55-100: 14\%$\end{tabular} &
  \begin{tabular}[c]{@{}c@{}}Female: $52\%$\\ Male: $48\%$\end{tabular} \\ \midrule
  
New England & $6934$ & $43$ & \begin{tabular}[c]{@{}c@{}}$9-16: 20\%$\\ $17-28: 32\%$\\ $29-41: 31\%$\\ $\textbf{42-54: 9\%}$\\ $\textbf{55-100: 8\%}$\end{tabular} &
  \begin{tabular}[c]{@{}c@{}}Female: $67\%$\\ Male: $33\%$\end{tabular} \\ \midrule
  
Mid Atlantic & $18550$ & $111$ & \begin{tabular}[c]{@{}c@{}}$9-16: 17\%$\\ $17-28: 23\%$\\ $29-41: 26\%$\\ $42-54: 19\%$\\ $55-100: 15\%$\end{tabular} &
  \begin{tabular}[c]{@{}c@{}}Female: $57\%$\\ Male: $43\%$\end{tabular} \\ \midrule
  
Midland & $12554$ & $73$ & \begin{tabular}[c]{@{}c@{}}$9-16: 16\%$\\ $17-28: 26\%$\\ $29-41: 31\%$\\ $42-54: 16\%$\\ $55-100: 11\%$\end{tabular} &
  \begin{tabular}[c]{@{}c@{}}Female: $63\%$\\ Male: $37\%$\end{tabular} \\ \midrule
  
LatinX & $9408$ & $56$ & \begin{tabular}[c]{@{}c@{}}$9-16: 16\%$\\ $17-28: 36\%$\\ $29-41: 36\%$\\ $\textbf{42-54: 8\%}$\\ $\textbf{55-100: 4\%}$\end{tabular} &
  \begin{tabular}[c]{@{}c@{}}Female: $66\%$\\ Male: $34\%$\end{tabular} \\ \midrule
  
Asian & $5228$ & $32$ & \begin{tabular}[c]{@{}c@{}}$9-16: 23\%$\\ $17-28: 36\%$\\ $29-41: 35\%$\\ $\textbf{42-54: 4\%}$\\ $\textbf{55-100: 2\%}$\end{tabular} &
  \begin{tabular}[c]{@{}c@{}}Female: $62\%$\\ Male: $38\%$\end{tabular} \\ \bottomrule
\end{tabular}
}
\caption{Statistical distribution of audio samples for each dialectal region in terms of age and gender in the test split. In bold are the categories for which we have less than $10\%$ of data points.}
\label{tab:confusion-bias}
\end{table}

\subsection{Ethical considerations on the audio collection}

To perform the audio collection, we commissioned a third-party data collection company to hire $1,080$ American English speakers (initially, later reduced to $1,039$ for quality reasons) who met the set of criteria we defined on ethnicity, age, gender, and dialectal regions, based on self declaration. 
These criteria amount to $80$ distinct demographic groups (not considering the smaller subset with known ethnicity), such as the 9-16 year old LatinX female speakers group. We further imposed a minimum number of speakers in each of these groups and a minimum amount of audio recording per speaker to guarantee a balanced distribution and enough power for the statistical tests. We provided recording scripts, as defined in \ref{sec:transcripts}.

The demographic category for a speaker was based on self declaration and no other personal information was collected apart from gender, age, dialectal region and ethnicity. We also explicitly collected consent from the speakers to re-distribute audio recordings of their voice for non-commercial academic and research purposes. The way the data was collected and will be distributed is fully compliant with the GDPR.

The third-party data collection company informed us that all participants were paid well above minimum wage. Hourly rate ranges from $\$15$ to $\$40$ per hour depending on the task requirements, such as the task duration.  The estimated time of completion was tracked and adjusted based on previous collections and internal testing of each task to reflect the median time of completion. 

All participant recruitment for child recordings is targeted towards parents who worked with their child(ren) to complete the recordings. The privacy policy of the third-party data collection company specifies that they do not process any personal data of children under 16 years of age without consent given or authorized by the holder of parental responsibility over the child. 
The payment is  done directly to the parent of the child.

\section{Bias Assessment Methodology} 
\label{sec:methodology}

\subsection{Approach}

In this section, we present the proposed methodology to assess the demographic biases of any given ASR system in the specific context of a music voice assistant.
The SLU metric we choose as proxy for user experience is \textbf{Exact Match} (EM, also called utterance-level accuracy in some works \citet{kashiwagi2023tensor}). An utterance $i$ is said to be exactly parsed ($EM_i=1$) if both the correct intent and all the correct slots are retrieved (else $EM_i=0$). In the case of a music voice assistant, EM is particularly relevant due to the intrinsic complexity of the slots values, especially artists or songs names that may be cross-lingual (e.g. Spanish songs or artists in an English request) and might be harder to pronounce for some demographic groups.

We propose to conduct statistical tests with logistic regressions \citet{mccullagh2019generalized} in order to assess the presence of demographic bias within the SLU system, i.e. to show whether the observed performance disparities are significant. EM serves as \textbf{binary response variable} for the tests. The subpopulations of interest are described by \textbf{categorical explanatory variables}: gender, age, dialectal region, and ethnicity.

Statistical testing with logistic regression helps to identify the effect of a single variable (e.g. singling out the effect of gender), via \textbf{univariate} models (Section~\ref{sec:univariate}) on the one hand. It also allows for identification of mixed effects (e.g. dialectal region \textit{and} gender) through \textbf{multivariate} approaches (Section~\ref{sec:multivariate}) on the other hand, as there could be demographic correlation between variables. In any case, a descriptive analysis of the data is always needed for the interpretation of such tests. 

The proposed methodology is valid under several conditions. First, we assume that each observation is independent, i.e. we do not take into account the speaker-level effect. We may therefore operate at the utterance-level for clarity of the interpretation and results. Second, we suppose that the speakers in our dataset are representative of their sub-group. Third, we expect the response variable (EM) to be a function of a linear combination of the considered explanatory variables (age, gender, dialectal region, ethnicity). Finally, we presume that the transcripts' difficulty is uniformly distributed among speakers (i.e. there are no speakers who have more complicated utterances to pronounce than others).

\subsection{Statistical Analysis}

\subsubsection{Univariate models}
\label{sec:univariate}
 
Logistic regression is a powerful statistical tool to measure demographic bias: it can shed light on the magnitude of the bias on the response variable through the estimated coefficients (fitted via maximum likelihood) and odds ratios (ORs). 
The probability that the Exact Match for observation $i$ is $1$: $\pi_i=\mathds{P}(EM_i=1)$ can be estimated by taking a monotonic real function $g(.): [0,1] \rightarrow \mathds{R}$ such that $g(\pi_i)=\beta_0 + \beta_1 x_i$, for instance the \texttt{logit} function: 
\begin{equation}    
g(\pi_i)=\ln\frac{\pi_i}{1-\pi_i}\\
\text{ with }g^{-1}(y)=\frac{e^y}{1+e^y} ,
\end{equation}
and $x_i$ the value of the explanatory variable for observation $i$, $\beta_1$ its associated coefficient and $\beta_0$ a constant. The corresponding odds is:
\begin{equation}
    \Omega_i(x_i) = \frac{\mathds{P}(EM_i=1)}{\mathds{P}(EM_i=0)} = \frac{\pi_i}{1 - \pi_i} = \exp(\beta_0 + \beta_1 x_i),
\end{equation}
i.e. given $x_i$, it is the relative chance of having $\mathds{P}(EM_i=1)$ compared to $\mathds{P}(EM_i=0)$.

Illustrating this on the binary explanatory variable \textit{gender}, with $x_i=\mathds{1}_{\text{male}}$, the OR is defined as:
\begin{equation}
    OR= \frac{\Omega_i(\mathds{1}_{x_i})}{\Omega_i(\mathds{1}_{1-x_i})} = \frac{\exp{(\beta_0 + \beta_1)}}{\exp{\beta_0}} = \exp{\beta_1},
\end{equation}
i.e. $\exp{\beta_1}$ represents the odds of a male speaker to be exactly parsed compared to a female one, all other things being equal. If $OR=1$, then \textit{gender} has no impact on $\mathds{P}(EM_i=1)$ (independence).


Additionally, we perform hypothesis testing (Wald test \citet{wald1943tests}) on the estimated parameter $\beta_1$ where the null is $H_0: \beta_1=0$. The latter is rejected at the $\alpha$-level if the corresponding p-value is lower than $\alpha$. 
This can be easily extended to non-binary categorical variables by following the method described in \ref{sec:multivariate}. It is used to test if all coefficients but the constant are null. Rejecting this null hypothesis means that at least one estimated coefficient is not null, but it does not inform us on which one it is. To get a better grasp at which modality impacts the \textit{EM} with respect to the reference modality, one must look at the individual p-values of each coefficient (or, equivalently, at the confidence intervals of the ORs). Note that every coefficient must be interpreted with respect to the reference group (e.g. \textit{female} in the above example).

\subsubsection{Multivariate models}
\label{sec:multivariate}

As mentioned previously, there might be confounding variables that could lead to spurious correlations and impact the statistical validity of our analysis. Gender is often described as such in the literature \citet{tatman2017gender} and age can be another one. To uncover such potential issues, we perform adjustment tests by augmenting the previous univariate models. For the sake of simplicity, let's illustrate the approach with two discrete variables: gender ($g=\mathds{1}_{\text{male}}$) and dialectal region ($d_j=\mathds{1}_{\text{j}}$, with $J=6$ modalities, leaving out the reference group). For the observation $i$, the multivariate model writes:
\begin{equation}
    g(\pi_i) = \beta_0 + \sum_{j=1}^{J} \beta_{1j} d_{ij} + \beta_2 g_i.
\label{eq:multivariate}
\end{equation}

To assert whether gender could be a potential confusion variable with respect to the dialectal region (i.e. are observed differences in EMs between dialectal regions due to gender imbalance?), we compare model (\ref{eq:multivariate}) to the univariate one:
\begin{equation}
    g(\pi_i) = \beta^\prime_0 + \sum_{j=1}^{J} \beta^\prime_{1j} d_i.
\label{eq:univariate}  
\end{equation}

Our proposed procedure to assess whether adding gender to the model adds significance, i.e. $H_0: \beta_2=0$, is as follows:
\begin{itemize}
    \item Maximum Likelihood Estimation (MLE) of $(\ref{eq:multivariate})$ and $(\ref{eq:univariate})$.
    \item Log-likelihood ratio test (LLR). $L_u$ is the log-likelihood (natural logarithm of the MLE function) of (\ref{eq:univariate}) and $L_m$ the one of (\ref{eq:multivariate}). Under $H_0$, the test statistic $T=2(L_m - L_u) \sim \chi^2_{r}$ where $r$ is the difference between the number of parameters of the two models. Similarly as before, $H_0$ can be rejected at the $\alpha$-level when $T>q_{r,1-\alpha}$, where $T$ is the test statistic and $q_{r,1-\alpha}$ the $1-\alpha$ quantile of a $\chi^2$ distribution with $r$ degrees of freedom.
\end{itemize}

Rejecting $H_0$ implies that gender adds information to the model. To further assess whether it confounds the effect of the dialectal region, we compare the p-values of the coefficients associated to the dialectal regions in (\ref{eq:multivariate}) and (\ref{eq:univariate}). If the conclusions of the Wald tests in (\ref{eq:univariate}) are changed (i.e. p-values becoming lower or greater than the $\alpha$ level), we can say that gender is a confounding factor for dialectal region, else there is no statistical evidence supporting this claim.


\section{Bias Assessment Experiments}
\label{sec:experiments}

In the following we propose an illustration of the proposed statistical method for bias assessment on the SLU task. To do so, we take an off-the-shelf ASR model (\textbf{wav2vec2.0} from~\citet{baevski2020wav2vec}) that we fine-tune on part of the train split of Sonos Voice Control bias assessment dataset to get audio transcription. Intent Classification (IC), Slot Filling (SF) and computation of the EM metric are done by using a \textbf{JointBERT} SLU model~\citet{chen2019bert}. The goal of this Section is to demonstrate the capabilities of our proposed dataset and methodology to quantify demographic bias in voice assistants. Consequently, the ASR and SLU models have not been particularly optimized for bias mitigation.

\subsection{Models}

\textbf{ASR model description.} The chosen ASR is an E2E Wav2Vec2-Large-960h model, developed by Meta and pre-trained using self-supervised learning (SSL) on $960$ hours of English speech data from the Librispeech dataset introduced by~\citetlanguageresource{panayotov2015librispeech} and fine-tuned on a subset of the train split of the Sonos Voice Control bias assessment dataset: audio samples from speaker groups 2 and 3 ($192$ speakers). It represents $30,602$ audio samples 
which amounts to $30.4$ hours of data. The entire dev split is used for validation.

In addition to the large wav2vec2.0 model, we incorporate an extra layer with $1024$ neurons and LeakyReLU as the activation function followed by a fully-connected layer and a final $40$-dimensional softmax layer, each dimension corresponding to a character. This neural network architecture comprises a total of $316.5$M trainable parameters. The weights of the two added layers were randomly initialized, while the weights of the wav2vec2.0 were initialized using the pre-trained weights. 

This SSL model is then fine-tuned with the additional layers, with a batch size of $12$, distributed across 4 NVIDIA V100 32GB GPU cards. Two optimizers are used: Adadelta~\citet{zeiler2012adadelta} for updating the additional layers' weights and Adam~\citet{kingma2014adam} for fine-tuning the SSL model, with initial learning rates of $1.0$ and $\num{1e-4}$ respectively. The maximum number of epochs is set to $80$: the best model on the validation is obtained at epoch~$72$, with a word error rate of $4.45\%$ on the dev split of the Sonos Voice Control bias assessment dataset. Processing an entire epoch takes around 52 minutes. During the fine-tuning process, a SpecAugment data augmentation technique was applied to the audio signal~\citet{park2019specaugment}.
The ASR system has been implemented by using the open source Speechbrain toolkit~\citet{ravanelli2021speechbrain}.

\textbf{SLU model description.} The SLU model is the \texttt{bert-base-uncased} version of the JointBERT~\citet{chen2019bert} model, pre-trained on BookCorpus ($800$M words)~\citet{Zhu2015AligningBA} and English Wikipedia ($2,500$M words). One linear layer is added on top of BERT and the IC weight in the cross-entropy loss is set to $0.01$. The batch size is $128$ and the transcripts are sorted by length and grouped into length buckets of size $10$ in order to ensure that batches are not padded too much. The whole network ($109$M parameters) is fine-tuned on the transcripts and labels of the full train split of the Sonos Voice Control dataset. During training, each transcript has been augmented once by replacing slot values by randomly drawing from all possible values in the train split. We use the Adam optimizer with a learning rate of $\num{5e-5}$; no dropout is applied. We train the model for $20$ epochs and select the checkpoint achieving the smallest weighted cross-entropy loss on the dev set as final model.

\subsection{Results}
\label{sec:results}
\begin{figure*}[!htbp]
    \centering
    \resizebox{0.95\textwidth}{!}{
    \renewcommand{\arraystretch}{.5}
     \begin{tabular}{cc}
        \centering
         \includegraphics[width=0.8\textwidth]{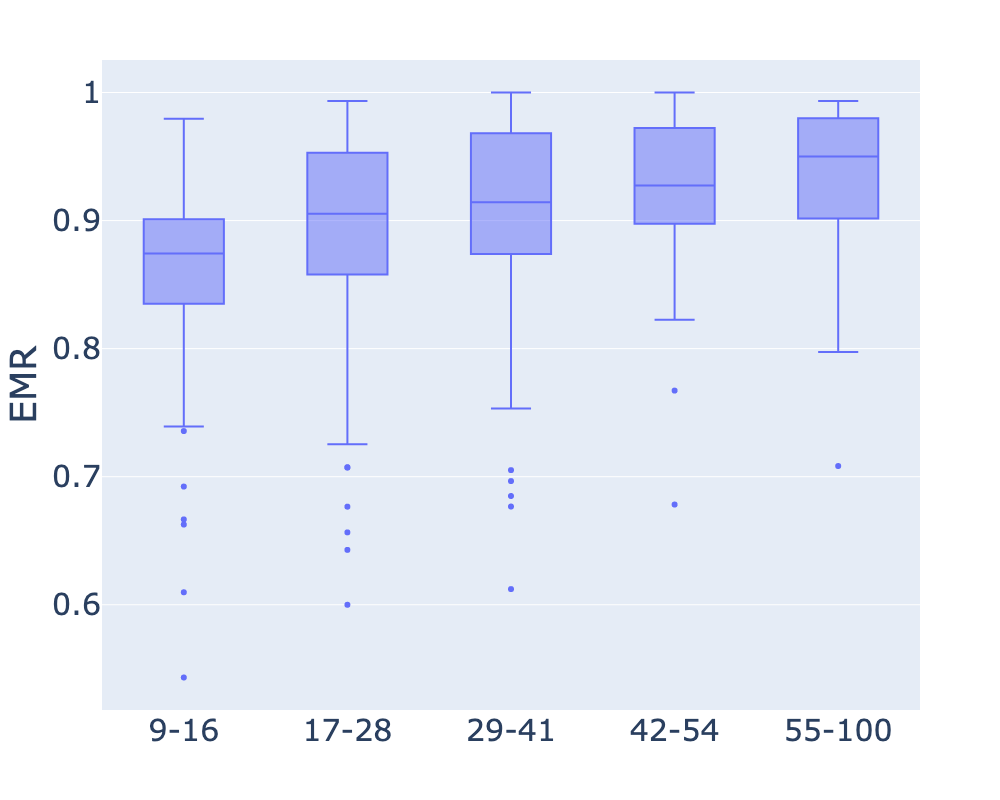} 
         &  
         \includegraphics[width=0.8\textwidth]{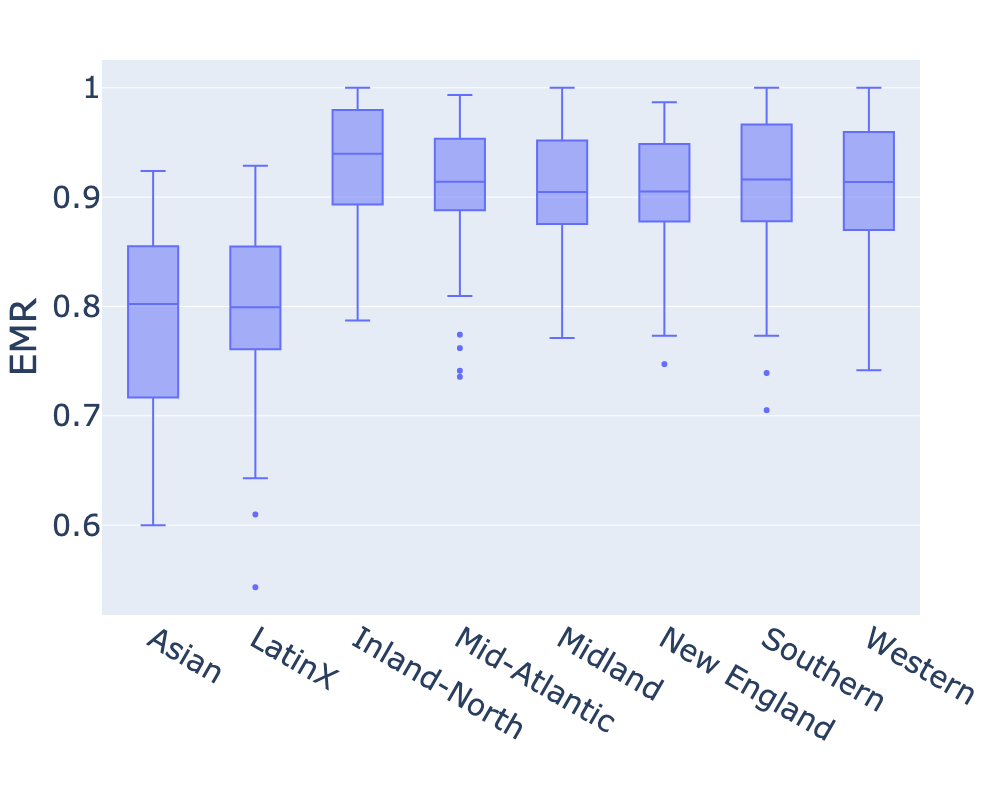} \\
         \large (a) Age group & \large (b) Dialectal region\\
         \includegraphics[width=0.8\textwidth]{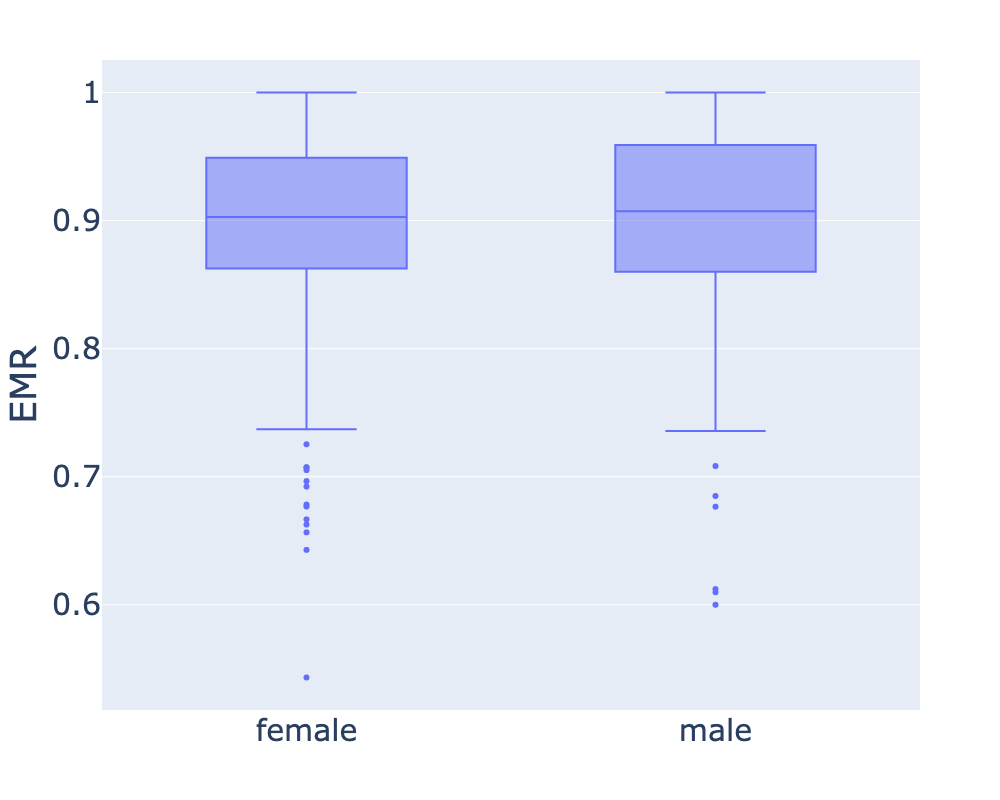} &
         \includegraphics[width=0.8\textwidth]{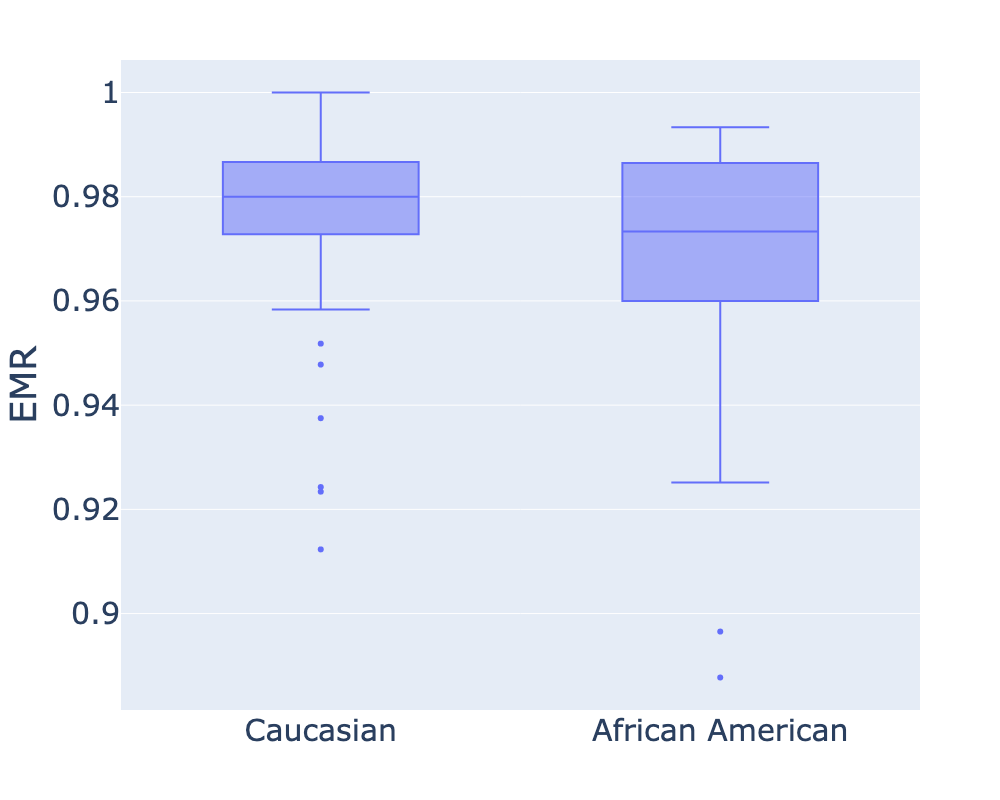} \\
         \large (c) Gender & \large (d) Ethnicity
    \end{tabular}}
    \caption{Exact Match Ratio (EMR) per speaker's demographic group. Points indicate individual speakers.}
    \label{fig:emr-w2v}
\end{figure*}

We compute the Exact Match Ratio (EMR, the fraction of exactly parsed audio samples) on the test split, using the fine-tuned E2E model for audio transcription and the fine-tuned SLU model to retrieve intents and slots. We find an EMR of $89\%$ ($76\%$ for \texttt{PlayMusic} and $99\%$ for \texttt{TransportControl}) and a WER of $2.5\%$. The variation of EMR per demographic group can be found in Figure~\ref{fig:emr-w2v}. Some individual speakers (outlier points) are much less well understood than others in the same demographic group. This is consistent with the literature~\cite{tatman2017effects, tatman2020sociolinguistic, feng2023towards}. Consideration on speaker variability and more details on WER can be found in~Appendix~\ref{sec:appendix-statistical-analysis}.

Applying the method described in Section~\ref{sec:methodology}, we find statistically significant evidence of bias for all considered explanatory variables in the \textbf{univariate} setting at the $5\%$ level. Regarding \textbf{gender}, men are significantly better recognized than women ($OR=1.05$, $p=0.017$). However, the OR is very close to $1$ and we see in panel (c) of Figure~\ref{fig:emr-w2v} that the difference is slim.

Tests on the \textbf{age} variable show that children are not recognized as well ($OR=0.85$, $p=\num{4.145e-8}$ for the 9-16 group) as younger adults (17-28yo, reference group), while older groups are increasingly better recognized (ORs are $1.11$, $1.48$, and $1.70$ for 29-41, 42-54, and 55-100 respectively, all p-values are $<\num{1e-4}$). Interestingly, we note that when isolating \texttt{Transport Control} requests only (simpler patterns, without music entities; see Section~\ref{sec:transcripts} for the exact definition), the disparities observed between children and younger adults are not statistically significant anymore ($p=0.070$).

For \textbf{dialectal region}, every group is better recognized than the Asian reference group. However, while requests from all American regional groups have around 3 times more chance to be exactly parsed than Asians' (ORs are around $3$, all p-values are $<\num{1e-89}$), this is only just slightly the case for the LatinX group ($OR=1.15$). This is clearly visible in panel (b) of Figure~\ref{fig:emr-w2v}. On the smaller \textbf{ethnicity} dataset, we found that Caucasian speakers are better understood than African Americans ($OR=1.59$, $p=\num{5.5e-5}$). Additional univariate tests were performed; results can be found in Appendix~\ref{subsec:appendix-uni}.

The \textbf{multivariate} analysis introduced in Section~\ref{sec:multivariate} is key to shine light on possible mixed effects. We detect several confusion variables. On one hand, \textbf{dialect is a confounding factor for gender}: the LLR test is statistically significant ($T=1748>q_{7,0.05}=14.07$) and the coefficient associated to \textit{male} is no longer significant ($p=0.8>0.05$). The difference observed in the univariate case for gender was actually due to the dialectal region. This hypothesis is confirmed by the gender distribution of the Asian group found in Table~\ref{tab:confusion-bias} (and to a lesser extent of the LatinX group): it is more skewed towards female speakers, while being also less well recognized than the other groups.

\begin{figure}[h!]
\begin{center}
    \includegraphics[width=1.05\columnwidth]{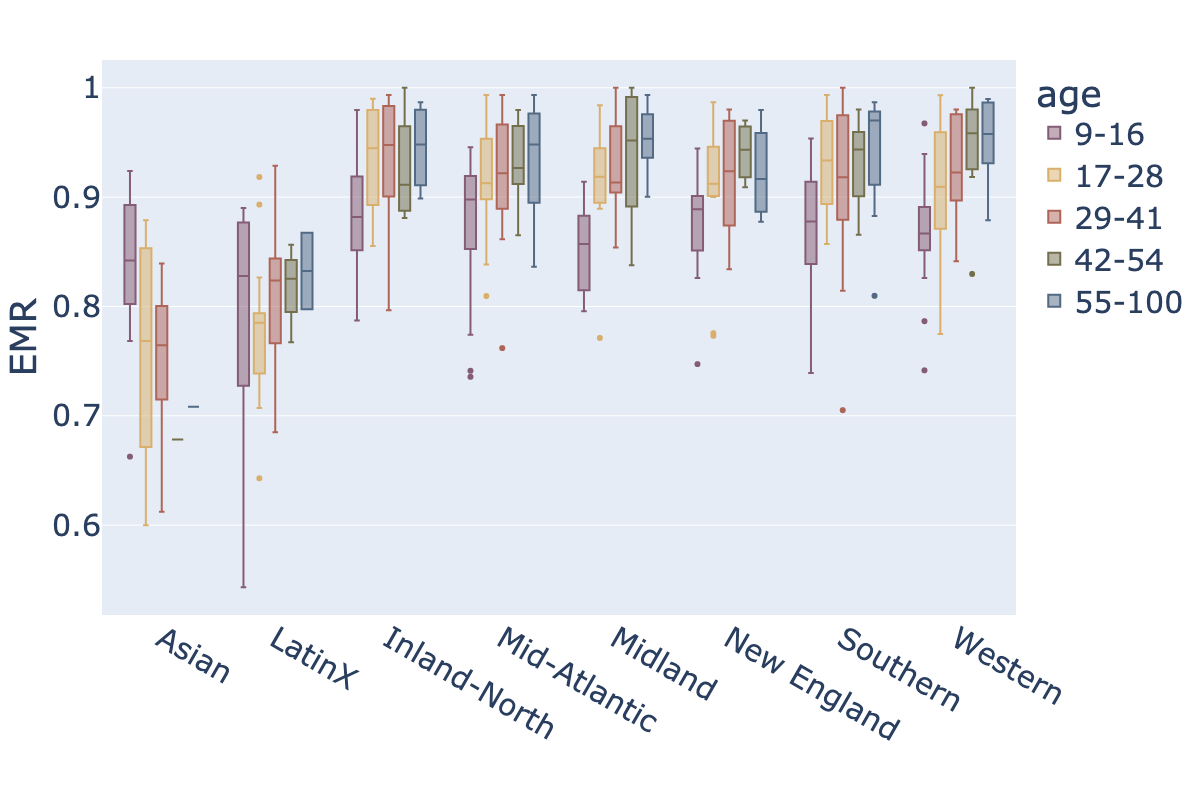}
\caption{Interaction effect on the Exact Match Ratio (EMR) of age and dialectal region. By splitting across dialectal regions, the differences in EMR between age groups are getting wider compared to Figure \ref{fig:emr-w2v}(a).}
\label{fig:boxplot-splitted-dialectal-region-age}
\end{center}
\end{figure}

On the other hand, combining dialectal region and age brings more significance to the model as in both cases (adjustments of dialectal region on age and of age on dialectal region), $H_0$ is rejected (respectively, $T=193>q_{4,0.05}=9.49$ and $T=1559>q_{7,0.05}=14.07$). The conclusions of the respective univariate tests are not changed (Wald p-values are still lower than $0.05$). However a \textbf{cross-effect of age and dialectal region} is brought to light since the corresponding ORs decrease from the univariate to the multivariate case (for instance, from $3.12$ to $2.91$ for the Mid-Atlantic group). This means that some of the difference observed between dialectal regions is actually due to the age disparity (and vice versa). This effect can be seen in Figure~\ref{fig:boxplot-splitted-dialectal-region-age}. Interestingly, younger Asian and LatinX speakers are better understood while it is the opposite for American regional groups.

By restricting our multivariate analysis to the subset of observations for which we have the ethnicity label, we did not find evidence of any confusion bias between ethnicity and the $3$ other demographic groups. Appendix~\ref{subsec:appendix-multi} displays the exhaustive and systematic analysis of all multivariate tests.

\section{Conclusion and Discussion}

This paper introduces the Sonos Voice Control bias assessment dataset, an open corpus made of $170,737$ audio samples ($166$ hours) from $1,039$ different speakers with their transcripts, labels and demographic tags (age, gender, dialectal region and ethnicity). We also propose a statistical bias assessment methodology (open sourcing its implementation), at the univariate and multivariate levels, tailored to the specific context of voice assistants. We consider Spoken Language Understanding (SLU) metrics measuring the interpretability of a user's request, rather than the standard transcription accuracy, as we believe it is more representative of end-to-end user experience. After describing the dataset (Section~\ref{sec:dataset}) and the statistical methodology (Section~\ref{sec:methodology}), their capabilities are illustrated with state-of-the-art ASR and SLU models (Section~\ref{sec:experiments}). Results on this example show statistically significant disparities in terms of SLU metrics across age, dialectal region and ethnicity. Second-order considerations allow to unveil mixed effects between dialectal region, gender and age. We hope that releasing this dataset and statistical methodology will foster research on demographic bias for voice assistants.

We identify several limitations in this study. Descriptive analysis of the dataset shows that the empirical distribution of a demographic variable of interest is not the same for the different modalities of another demographic variable of interest (e.g. distribution of gender across dialectal regions; Table \ref{tab:confusion-bias}). These differences are statistically significant (unveiled by non-parametric independence tests; results are not presented here). Therefore there is a selection bias in our dataset leading to a confusion bias when analyzing data. It has to be taken into account for results’ interpretation and motivates the need to perform multivariate analysis in addition to univariate tests (as seen with gender in Section~\ref{sec:results}). 

A side effect of controlling the request distribution in the dataset, so that they are representative of a voice assistant in the music domain, is that the recorded samples are not spontaneous, but must follow an imposed transcript. This setting creates two main limitations in the dataset that might amplify bias against some demographic groups. First, evidence shows that the performance of ASR systems degrades in spontaneous speech conditions~\citet{nakamura2008differences} and this degradation may not be uniform across all groups. Second, imposing the transcript necessarily erases possible lexical and grammatical particularities of each population, which might also amplify bias. 

However, recent research seems to suggest that observed disparities in speech recognition accuracy rather stem from the acoustic model rather than the language model~\citet{koenecke2020racial}. Moreover, the music voice assistant use case is constrained to a small number of possible phrasings (think about how one can ask about a song, an artist or a volume set). This dampens the effects of variability in wording. Moreover, while reading is not a perfect simulation for speech that is directed at voice assistants, it allows for controlled collection and is the only feasible option for large-scale data collection such as ours.

We also want to emphasize that the demographic description of the speakers in the dataset has evidently strong limitations in terms of linguistic, social and geopolitical representation, and is only acceptable in terms of the linguistic approximations required for the use case at hand. In addition, any regional split over dialectal variation can  be considered as arbitrary, since it generalises over partial demographic estimations and since dialectal variation can be viewed as a linguistic continuum rather than a set of contiguous spaces. We also only consider a binary view of gender while there are more fluid experiences. Other factors such as literacy, level of education, social, cultural and economical background \citet{dichristofano2022performance, Chan2022TrainingAT} could be taken into account, as well as possible speech impairments (e.g. dysarthria in~\citet{tu2016relationship, moro2019study} or cleft lip and palate \citet{schuster2006evaluation}).

Moreover, modeling explicitly the speaker effect in the statistical analysis is possible (as a random effect for instance as in~\citet{liu2022model}) but would increase the complexity of our approach without bringing massive improvements. We argue that the main goal of this paper is to propose a simple yet effective statistical methodology to assess the demographic bias of any given SLU system and provide clear interpretation of the results.

Future work includes considering narrower age ranges to study the effects of younger children without reading capabilities, or elderly adults whose speech patterns may deviate significantly from standard speech. On another topic, it will be extremely interesting to generate far-field and noisy versions of the dataset by simulating realistic acoustic conditions (through room impulse response and reverberation simulation for instance): the performance of ASR systems is indeed particularly degraded in these conditions, while they are typical of domestic voice assistant usage.

\section{Acknowledgements}
We would like to thank the team of linguists at Sonos for their invaluable expertise and guidance throughout this project, we are especially grateful to Claudia Garber (Sonos) for her contributions to the discussion section, the Appendix, and help reviewing the paper. Thomas Nigoghossian (Sonos) provided valuable insight with the JointBERT NLU training. We thank Dr. Odette Scharenborg (Delft University of Technology) for fruitful discussions and support.

\nocite{*}
\section{Bibliographical References}
\label{sec:reference}

\bibliographystyle{lrec_natbib}
\bibliography{lrec-coling2024-example}

\section{Language Resource References}
\label{lr:ref}
\bibliographystylelanguageresource{lrec-coling2024-natbib}
\bibliographylanguageresource{languageresource}

\clearpage

\appendix
\appendix

\section{Appendix A: Sonos Voice Control Bias Assessment Dataset, additional descriptive statistics}
\label{sec:appendix-svc-dataset-stats-desc}

\subsection{Dialectal regions definition}
\label{subsec:appendix-dialectal-regions}

Dialectal regions in America are defined based on phonemic/phonetic, lexical, and syntactic features.
Speakers from a certain region may use certain vocabulary or lexical items that are specific to their region (for example `gym shoes' or `sneaker' all refer to the same athletic footwear but will be heard with varying frequency depending on the geographic location). In the case of voice assistants, particularly within the music domain, lexical differences and regionalisms do not pose many issues given the short and straightforward nature of interactions. In a similar way, syntactic features (such as the use of 'done' as an auxiliary verb in Southern English to express the past tense: I done had enough) do not pose many issues to speech recognition within voice assistant domains.

Regional varieties are also reflected in specific phonetic phenomena, such as non-rhoticity (dropping of the /r/ consonant in all environments except before a vowel, sometimes heard in Boston or New York), PIN-PEN merger (the two vowel sounds /I/ and /E/ merge before nasal sounds to sound the same, found in standard Southern American dialect), or COT-CAUGHT merger (/\textipa{A}/ and /\textipa{O}/ vowels to sound alike, heard in most of the country). Vowel quality and, more generally, phenomena related to the sound system of English pose more issues to voice recognition, especially in the music domain. Depending on how speakers pronounce music values and general queries, recognition may vary.

Identifying dialectal regions for any area or language is always imprecise. While what is defined in our data is one way of interpreting American English, there are of course many different ways of dividing regionality and the features that exist within these groups. Within any dialectal region, the prominence of the specific features of that dialect will vary greatly. While some may be inclined to further divide into even more specific groups, having data that represents multiple ages, genders and demographics within each dialectal region was also of importance.

The definition of dialectal region was also impacted by the necessity for adequate representation from each category, while also adhering to other constraints such as time and budget. Creating more dialectal groups would potentially impede the velocity of data collection or skew sample sizes.

Selecting these dialectal regions also facilitated easy identification of a user within a group. Without being able to listen to or speak with individuals to `verify' their dialectal region selection, having no more than six dialectal groups facilitated data collection with straight-forward parameters based on geographic location.

The \textbf{Asian and LatinX} categories were defined by identifying other large speaker groups in the US that may interact with voice assistants. These two user groups were defined as follows:
home country must be in Latin America for LatinX, in Southern or Eastern Asia for Asian
native language must be Spanish for LatinX, any Asian language for Asian
The definition of the LatinX group is different from the definition one may use for Chicano English (also known as Mexican-American English or Spanish English). Chicano English is primarily spoken by Mexican Americans in south-western states whose first language is English. We were interested in a larger pool than only Mexican American speakers. The speakers in our LatinX group are native Spanish speakers from any Latin American country. However, many of the features in Chicano English are also exhibited in the LatinX participant's speech. For example, /\textipa{D}/ stopping, or the replacing of the -th- sound with /\textipa{d}/ in words like `there' (`dere') was still prevalent. The distinction between /\textipa{I}/ and /\textipa{i}/ in some speakers is lost, making words like `fit' and `feet' sound alike. 

General monophthongization, which may be directly due to the quality of vowels in Spanish, is another quality of Chicano English that was seen in the LatinX group. The motivation behind criteria based on language and country was again largely due to time constraints, as it can be difficult to find enough speakers when more constraints are applied to dialectal definitions. It also facilitated the identification of speakers - being able to identify based on home country and native language is easier for participants than asking if they think they speak a certain dialect of a language.
The Asian group included Southern and Eastern Asian countries, in order to account for large populations of both regions in the US. While the Asian group has much more linguistic diversity within the speakers themselves, no further division was implemented based on budgeting and time constraints. Looking back, if we were to reproduce this approach, Indian English should probably have been separated from the other Asian groups.

\clearpage

\subsection{Ontology description}
\label{subsec:appendix-ontology}

Among all splits, there are $9,040$ unique transcripts, $8,114$ for \texttt{PlayMusic} and $926$ for \texttt{Transport Control}. Table \ref{tab:list-nb-slot-values-per-slot-name} provide the number of unique slot values for each \textit{slot name}, e.g. there are $221$ different radio names. 

\begin{table}[H]
\centering
\resizebox{0.9\columnwidth}{!}{%
\begin{tabular}{cc}
\toprule
\textbf{Slot name} & \textbf{\# of unique values} \\ \midrule
abs\_volume & 13 \\ \midrule
activity & 8 \\ \midrule
album\_name & 1626 \\ \midrule
artist\_name & 1715 \\ \midrule
call\_sign & 404 \\ \midrule
container\_qualif\_after\_artist & 7 \\ \midrule
container\_qualif\_before\_artist & 9 \\ \midrule
container\_type & 5 \\ \midrule
content\_type & 16 \\ \midrule
destination\_group & 1 \\ \midrule
destination\_target & 21 \\ \midrule
except & 2 \\ \midrule
frequency & 230 \\ \midrule
genre & 272 \\ \midrule
implicit\_content & 1 \\ \midrule
instead & 2 \\ \midrule
library & 22 \\ \midrule
location & 66 \\ \midrule
mood & 2 \\ \midrule
only & 1 \\ \midrule
origin\_target & 12 \\ \midrule
personal\_container\_name & 58 \\ \midrule
playback\_mode & 8 \\ \midrule
playlist & 599 \\ \midrule
program & 86 \\ \midrule
provider & 12 \\ \midrule
radio\_name & 221 \\ \midrule
rel\_volume & 10 \\ \midrule
shuffle & 1 \\ \midrule
song\_name & 2643 \\ \midrule
target & 298 \\ \midrule
target\_exception & 32 \\ \midrule
too & 3 \\ \midrule
volume\_down\_subj & 8 \\ \midrule
volume\_obj & 111 \\ \midrule
volume\_set\_subj & 28 \\ \midrule
volume\_shift\_subj & 9 \\ \midrule
volume\_up\_subj & 6 \\ \bottomrule
\end{tabular}%
}
\caption{Number of unique slot values per slot name in the proposed dataset.}
\label{tab:list-nb-slot-values-per-slot-name}
\end{table}

\tablecaption{List of available slot names per intent in the proposed dataset.}
\label{tab:list-slots-names-per-intent}
\tablefirsthead{\toprule \textbf{Intent}&\multicolumn{1}{c}{\textbf{Corresponding slot names}} \\ \midrule}
\tablehead{\toprule
\textbf{Intent}&\multicolumn{1}{c}{\textbf{Corresponding slot names}}\\}
\tabletail{\midrule}
\tablelasttail{%
 \\ \bottomrule}
\begin{xtabular}{p{0.37\columnwidth}p{0.4\columnwidth}}
AddToLibrary &
  \begin{tabular}[l]{@{}l@{}} $\sbullet$ personal\_container\_name\\ $\sbullet$ provider\\ $\sbullet$ library\\ $\sbullet$ content\_type\\ $\sbullet$ target\end{tabular} \\ \midrule
ChangeMusic &
  \begin{tabular}[l]{@{}l@{}} $\sbullet$ provider\\ $\sbullet$ content\_type\\ $\sbullet$ target\end{tabular} \\ \midrule
ChangeTarget &
  \begin{tabular}[l]{@{}l@{}} $\sbullet$ instead\\ $\sbullet$ container\_type\\ $\sbullet$ origin\_target\\ $\sbullet$ destination\_target\end{tabular} \\ \midrule
CheckBattery &
  $\sbullet$ target \\ \midrule
FollowArtist &
  $\sbullet$ provider \\ \midrule
Forward &
  \begin{tabular}[l]{@{}l@{}} $\sbullet$ provider\\ $\sbullet$ target\end{tabular} \\ \midrule
GetInfos &
  \begin{tabular}[l]{@{}l@{}} $\sbullet$ provider\\ $\sbullet$ content\_type\\ $\sbullet$ target\end{tabular} \\ \midrule
GroupTargets &
  \begin{tabular}[l]{@{}l@{}} $\sbullet$ only\\ $\sbullet$ target\_exception\\ $\sbullet$ too\\ $\sbullet$ except\\ $\sbullet$ destination\_group\\ $\sbullet$ target\end{tabular} \\ \midrule
Like &
  \begin{tabular}[l]{@{}l@{}} $\sbullet$ provider\\ $\sbullet$ content\_type\\ $\sbullet$ target\end{tabular} \\ \midrule
Mute &
  \begin{tabular}[l]{@{}l@{}} $\sbullet$ container\_type\\ $\sbullet$ only\\ $\sbullet$ target\_exception\\ $\sbullet$ too\\ $\sbullet$ provider\\ $\sbullet$ except\\ $\sbullet$ target\end{tabular} \\ \midrule
NextSong &
  \begin{tabular}[l]{@{}l@{}} $\sbullet$ provider\\ $\sbullet$ container\_type\\ $\sbullet$ target\end{tabular} \\ \midrule
Pause &
  \begin{tabular}[l]{@{}l@{}} $\sbullet$ provider\\ $\sbullet$ container\_type\\ $\sbullet$ target\end{tabular} \\ \midrule
Play &
  \begin{tabular}[l]{@{}l@{}} $\sbullet$ container\_type\\ $\sbullet$ only\\ $\sbullet$ volume\_obj\\ $\sbullet$ target\_exception\\ $\sbullet$ too\\ $\sbullet$ volume\_down\_subj\\ $\sbullet$ provider\\ $\sbullet$ except\\ $\sbullet$ volume\_set\_subj\\ $\sbullet$ instead\\ $\sbullet$ target\end{tabular} \\ \midrule
PlayMusic &
  \begin{tabular}[l]{@{}l@{}} $\sbullet$ container\_type\\ $\sbullet$ only\\ $\sbullet$ container\_qualif\_after\_artist\\ $\sbullet$ genre\\ $\sbullet$ target\_exception\\ $\sbullet$ radio\_name\\ $\sbullet$ activity\\ $\sbullet$ call\_sign\\ $\sbullet$ program\\ $\sbullet$ artist\_name\\ $\sbullet$ frequency\\ $\sbullet$ container\_qualif\_before\_artist\\ $\sbullet$ except\\ $\sbullet$ implicit\_content\\ $\sbullet$ shuffle\\ $\sbullet$ song\_name\\ $\sbullet$ album\_name\\ $\sbullet$ target\\ $\sbullet$ playlist\\ $\sbullet$ playback\_mode\\ $\sbullet$ mood\\ $\sbullet$ too\\ $\sbullet$ location\\ $\sbullet$ volume\_obj\\ $\sbullet$ provider\\ $\sbullet$ volume\_set\_subj\\ $\sbullet$ library\\ $\sbullet$ instead\end{tabular} \\ \midrule
PreviousSong &
  \begin{tabular}[l]{@{}l@{}} $\sbullet$ provider\\ $\sbullet$ target\end{tabular} \\ \midrule
RemoveFromLibrary &
  \begin{tabular}[l{0.5\columnwidth}]{@{}l@{}} $\sbullet$ personal\_container\_name\\ $\sbullet$ provider\\ $\sbullet$ library\\ $\sbullet$ content\_type\\ $\sbullet$ target\end{tabular} \\ \midrule
Repeat &
  \begin{tabular}[l]{@{}l@{}} $\sbullet$ provider\\ $\sbullet$ content\_type\\ $\sbullet$ target\end{tabular} \\ \midrule
RestartSong &
  \begin{tabular}[l]{@{}l@{}} $\sbullet$ provider\\ $\sbullet$ container\_type\\ $\sbullet$ target\end{tabular} \\ \midrule
Resume &
  \begin{tabular}[l]{@{}l@{}}$\sbullet$ provider\\ $\sbullet$ container\_type\\ $\sbullet$ target\end{tabular} \\ \midrule
Rewind &
  \begin{tabular}[l]{@{}l@{}} $\sbullet$ provider\\ $\sbullet$ target\end{tabular} \\ \midrule
Shuffle &
  \begin{tabular}[l]{@{}l@{}} $\sbullet$ provider\\ $\sbullet$ target\end{tabular} \\ \midrule
Stop &
  \begin{tabular}[l]{@{}l@{}} $\sbullet$ container\_type\\ $\sbullet$ only\\ $\sbullet$ target\_exception\\ $\sbullet$ too\\ $\sbullet$ provider\\ $\sbullet$ except\\ $\sbullet$ target\end{tabular} \\ \midrule
StopAndStartTarget &
  \begin{tabular}[l]{@{}l@{}} $\sbullet$ instead\\ $\sbullet$ container\_type\\ $\sbullet$ origin\_target\\ $\sbullet$ destination\_target\end{tabular} \\ \midrule
StopMode &
  \begin{tabular}[l]{@{}l@{}} $\sbullet$ playback\_mode\\ $\sbullet$ provider\\ $\sbullet$ target\end{tabular} \\ \midrule
UnfollowArtist &
  $\sbullet$ provider \\ \midrule
UngroupTargets &
  \begin{tabular}[l]{@{}l@{}} $\sbullet$ only\\ $\sbullet$ target\_exception\\ $\sbullet$ too\\ $\sbullet$ except\\ $\sbullet$ target\end{tabular} \\ \midrule
Unlike &
  \begin{tabular}[l]{@{}l@{}} $\sbullet$ provider\\ $\sbullet$ content\_type\\ $\sbullet$ target\end{tabular} \\ \midrule
Unmute &
  \begin{tabular}[l]{@{}l@{}} $\sbullet$ container\_type\\ $\sbullet$ only\\ $\sbullet$ target\_exception\\ $\sbullet$ too\\ $\sbullet$ provider\\ $\sbullet$ except\\ $\sbullet$ target\end{tabular} \\ \midrule
VolumeDown &
  \begin{tabular}[l]{@{}l@{}} $\sbullet$ container\_type\\ $\sbullet$ only\\ $\sbullet$ abs\_volume\\ $\sbullet$ target\_exception\\ $\sbullet$ too\\ $\sbullet$ volume\_shift\_subj\\ $\sbullet$ rel\_volume\\ $\sbullet$ volume\_down\_subj\\ $\sbullet$ provider\\ $\sbullet$ except\\ $\sbullet$ volume\_set\_subj\\ $\sbullet$ target\end{tabular} \\ \midrule
VolumeSet &
  \begin{tabular}[l]{@{}l@{}}$\sbullet$ container\_type\\ $\sbullet$ only\\ $\sbullet$ volume\_obj\\ $\sbullet$ target\_exception\\ $\sbullet$ too\\ $\sbullet$ except\\ $\sbullet$ volume\_set\_subj\\ $\sbullet$ target\end{tabular} \\ \midrule
VolumeUp &
  \begin{tabular}[l]{@{}l@{}}$\sbullet$ container\_type\\ $\sbullet$ only\\ $\sbullet$ abs\_volume\\ $\sbullet$ volume\_up\_subj\\ $\sbullet$ target\_exception\\ $\sbullet$ too\\ $\sbullet$ volume\_shift\_subj\\ $\sbullet$rel\_volume\\ $\sbullet$ provider\\ $\sbullet$ except\\ $\sbullet$ volume\_set\_subj\\ $\sbullet$ target\end{tabular}
\end{xtabular}%

\vspace{0.5cm}
Table \ref{tab:list-slots-names-per-intent} provides the complete list of available slot names of each intent available in the SVC Bias Assessment Dataset.

\subsection{Additional descriptive statistics per split}

In this subsection of the appendix, we provide additional figures showing the distribution, in terms of both samples and speakers in each of the provided split in the Sonos Voice Control Bias Assessment Dataset. 

\subsubsection{Test split}

\begin{figure}[h!]
\begin{center}
\includegraphics[scale=0.2]{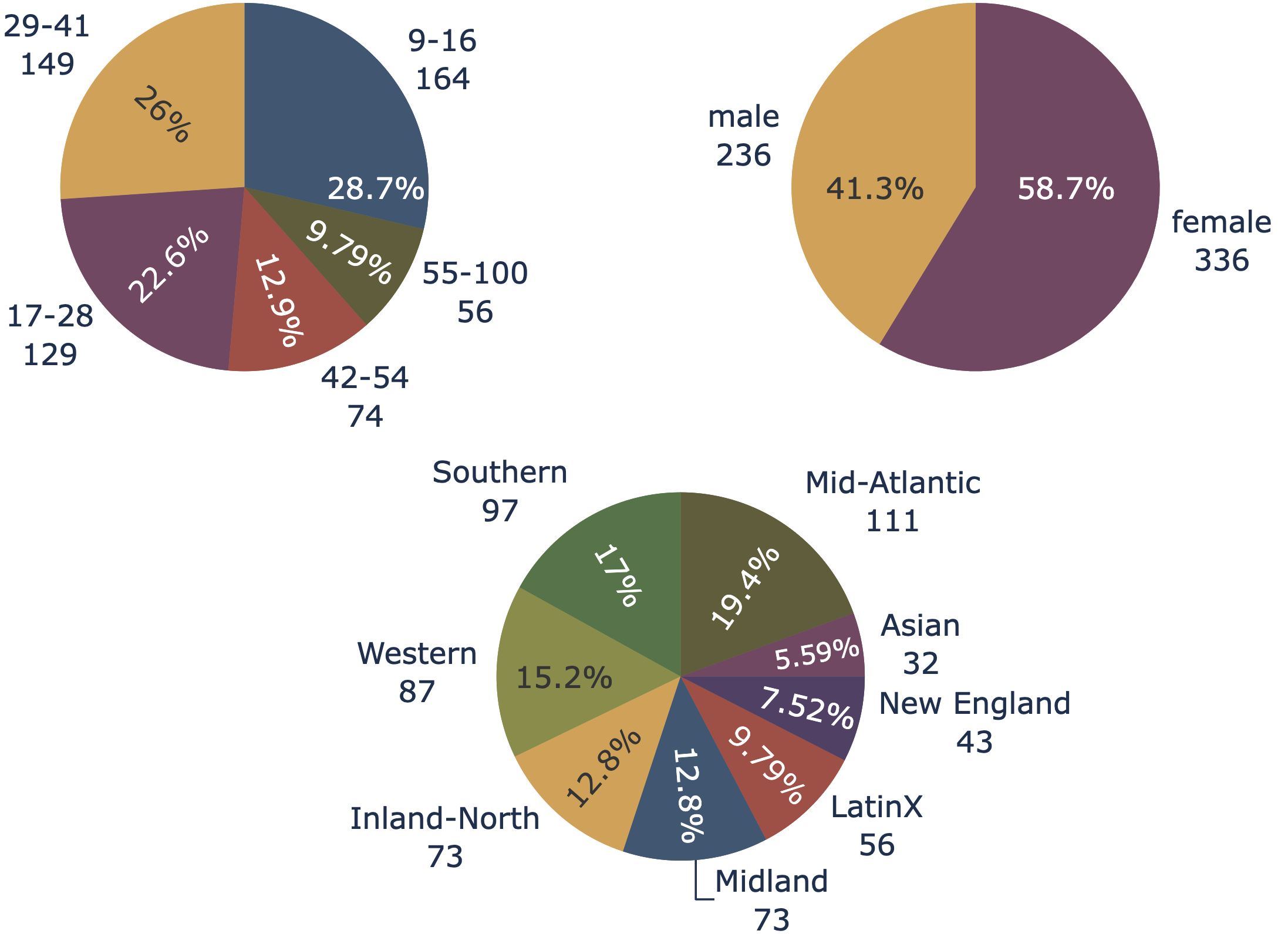}
\caption{Speaker distribution in the test split of the dataset in terms of age, gender, and dialectal region. The number of speakers in each group is displayed under the group label.}
\label{fig:speakers-distrib-test}
\end{center}
\end{figure}

Fig.~\ref{fig:speakers-distrib-test} displays the speaker distribution in the test split for each demographic group. 

\subsubsection{Train split}

\begin{figure}[h!]
\begin{center}
\includegraphics[scale=0.2]{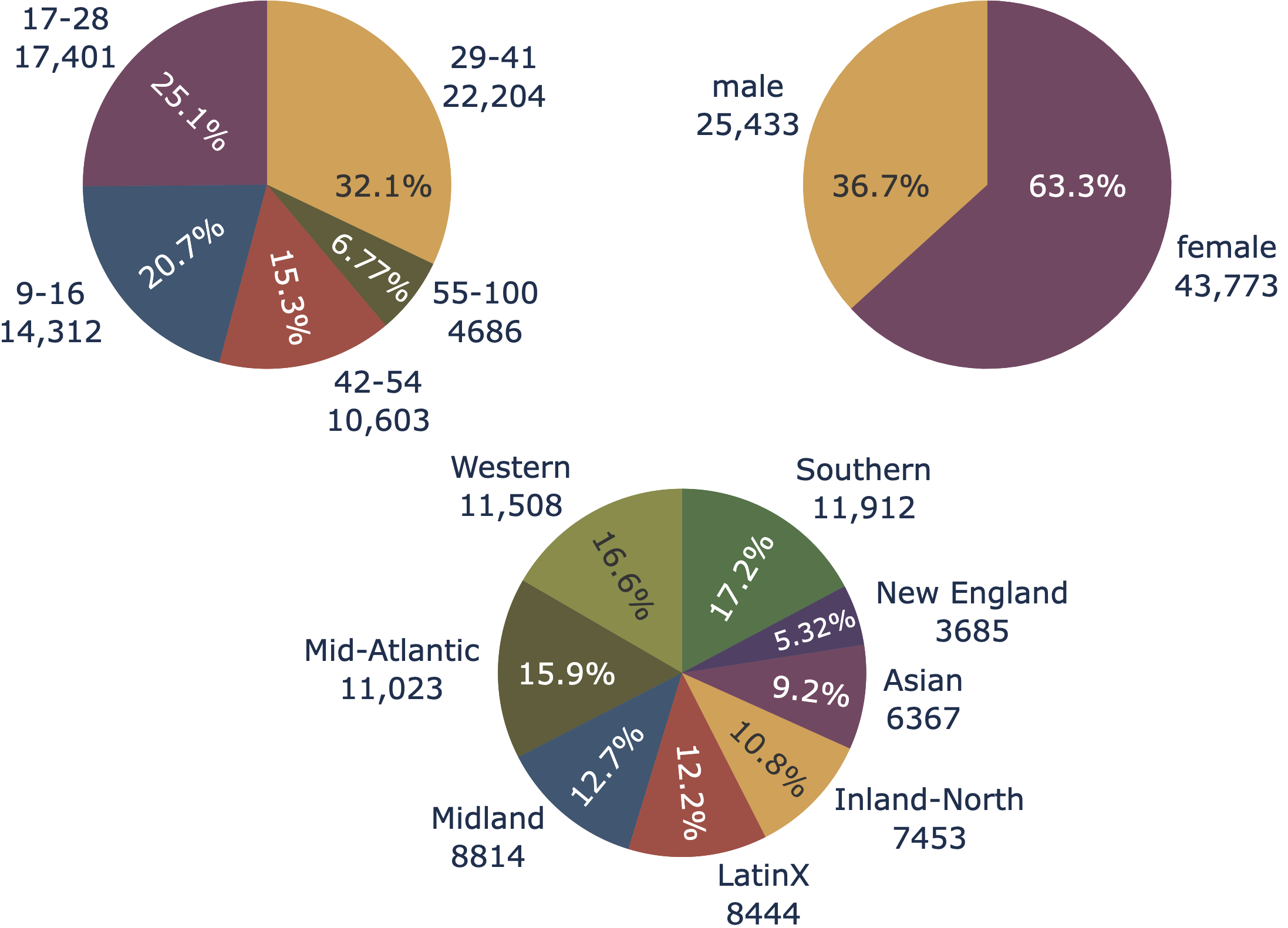}
\caption{Audio sample distribution in the train split of the dataset in terms of age, gender, and dialectal region. The number of samples in each group is displayed under the group label.}
\label{fig:samples-distrib-train}
\end{center}
\end{figure}

Fig.~\ref{fig:samples-distrib-train} displays the audio sample distribution in the train split for each demographic group. 

\begin{figure}[h!]
\begin{center}
\includegraphics[scale=0.2]{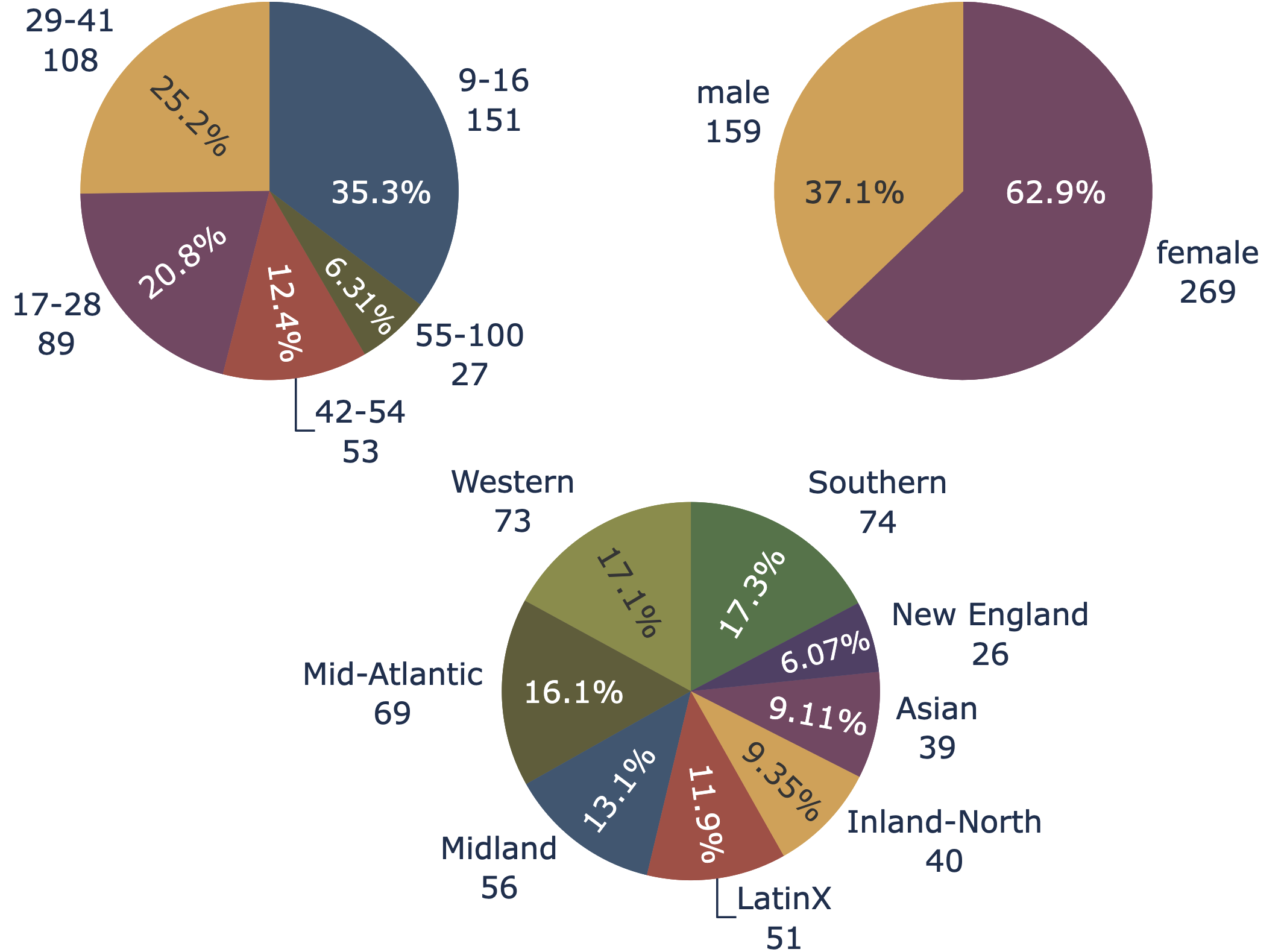}
\caption{Speaker distribution in the train split of the dataset in terms of age, gender, and dialectal region. The number of samples in each group is displayed under the group label.}
\label{fig:speakers-distrib-train}
\end{center}
\end{figure}

Fig.~\ref{fig:speakers-distrib-train} displays the speaker distribution in the test split for each demographic group. 

\subsubsection{Development split}

\begin{figure}[h!]
\begin{center}
\includegraphics[scale=0.2]{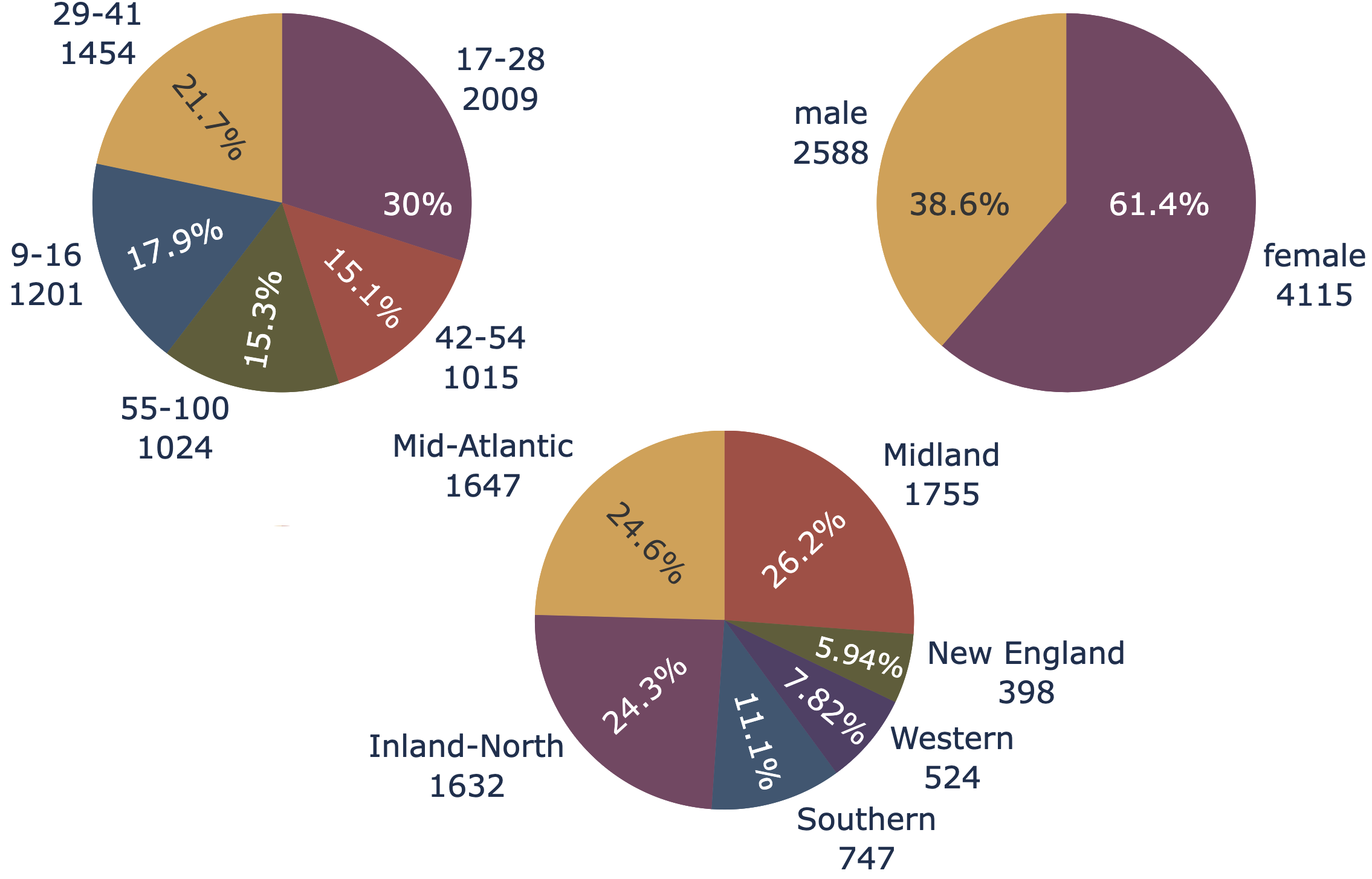}
\caption{Audio sample distribution in the development split of the dataset in terms of age, gender, and dialectal region. The number of samples in each group is displayed under the group label.}
\label{fig:samples-distrib-dev}
\end{center}
\end{figure}

Fig.~\ref{fig:samples-distrib-dev} displays the audio sample distribution in the development split for each demographic group. 

\begin{figure}[h!]
\begin{center}
\includegraphics[scale=0.2]{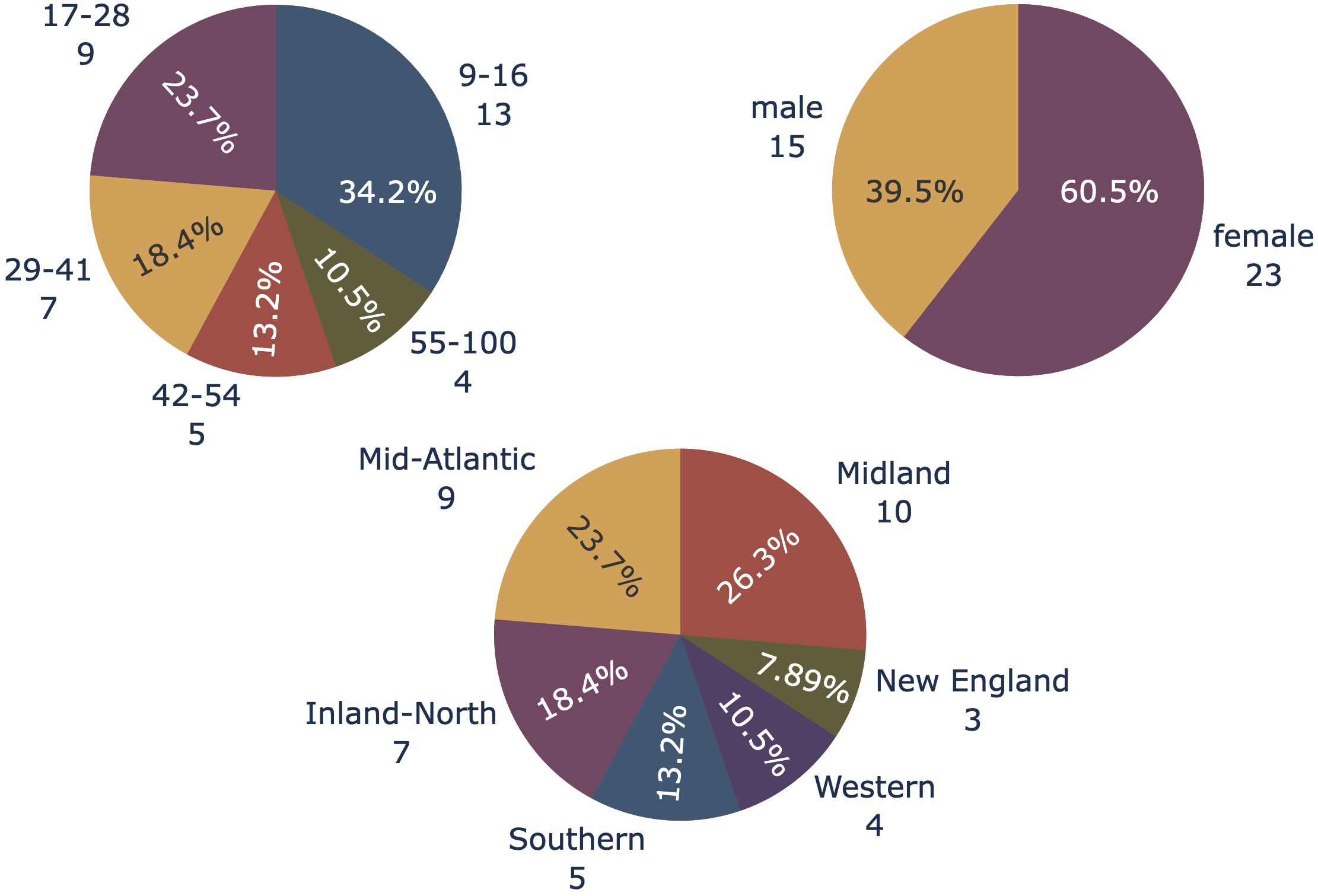}
\caption{Speaker distribution in the development split of the dataset in terms of age, gender, and dialectal region. The number of samples in each group is displayed under the group label.}
\label{fig:speakers-distrib-dev}
\end{center}
\end{figure}

Fig.~\ref{fig:speakers-distrib-dev} displays the speaker distribution in the development split for each demographic group. 

There is no non-native speaker (Asian or LatinX). This is primarily due to the small size of this set comprising only $38$ speakers.

\vspace{5cm}

\subsection{Ethnicity dataset}
\label{subsec:appendix-ethnicity}

Race and ethnicity are inherently difficult to define, as these words can mean different things to different people. There are also constraints on what can be asked or assumed of participants when working with third party providers. For these reasons, users were asked to self-identify and we operated under the assumption that a person of any racial or ethnic group would know their own identification better than us enforcing any strict parameters. It should also be noted that dialects such as African American Vernacular English (AAVE) are social dialects and are therefore not tied to geographical location. While not all members of an ethnic group, such as African American/Black, will exhibit features of the associated dialect, for example AAVE, this approach was agreed upon with the third party provider in charge of hiring speakers.

 The \textbf{ethnicity} tag was only reported in the second campaign that we launched for which $98$ speakers have been recruited: $50$ Caucasian and $48$ African American speakers. We refer to this smaller dataset, only present in the test split, as the \texttt{ethnicity dataset}.

\begin{figure}[h!]
\begin{center}
\includegraphics[scale=0.2]{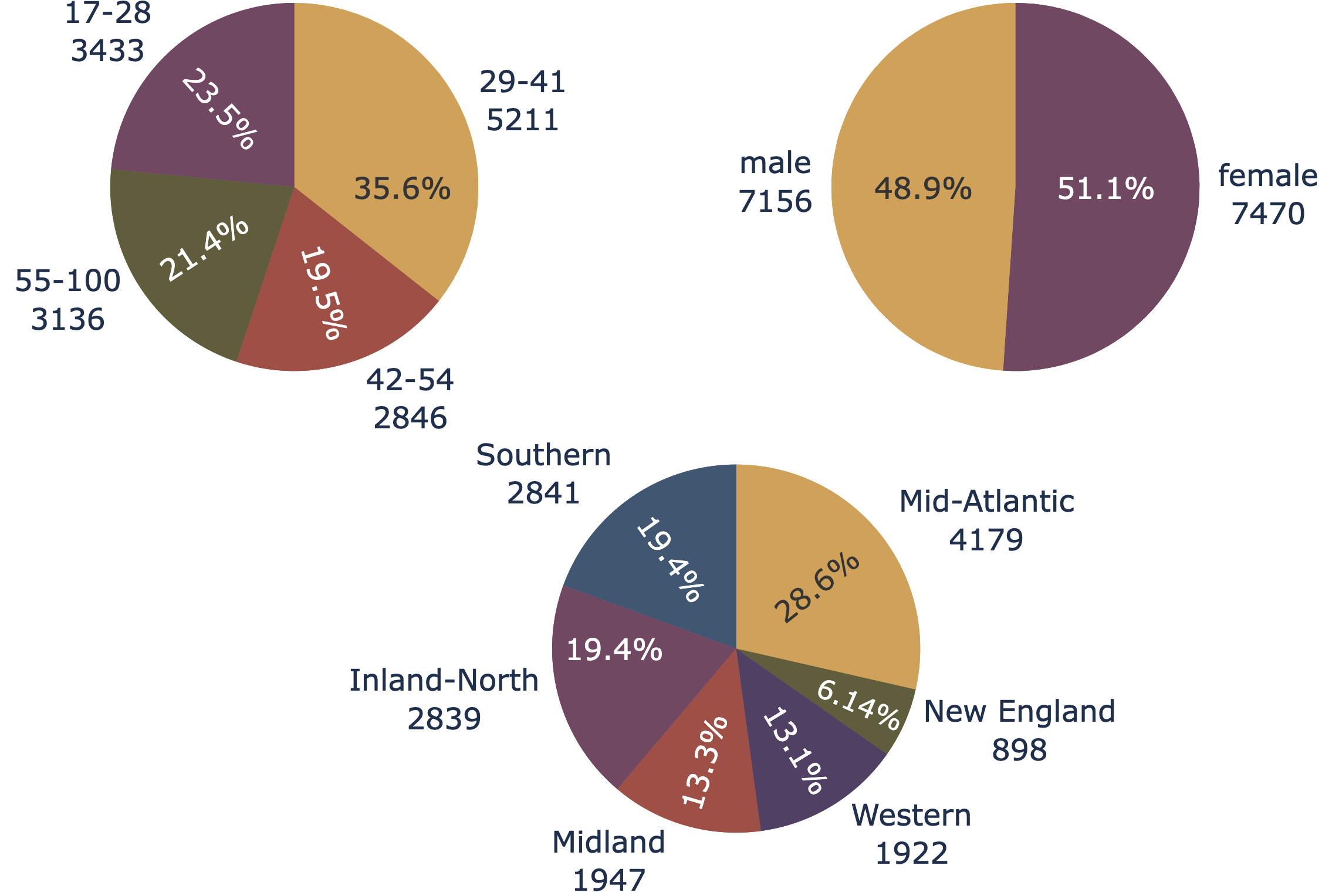}
\caption{Audio sample distribution in ethnicity dataset in terms of age, dialectal region and gender.}
\label{fig:audio-samples-distrib-ethnicity}
\end{center}
\end{figure}

Fig.~\ref{fig:audio-samples-distrib-ethnicity} displays the audio samples distribution in the ethnicity subset for each demographic group. 

\begin{figure}[h!]
\begin{center}
\includegraphics[scale=0.2]{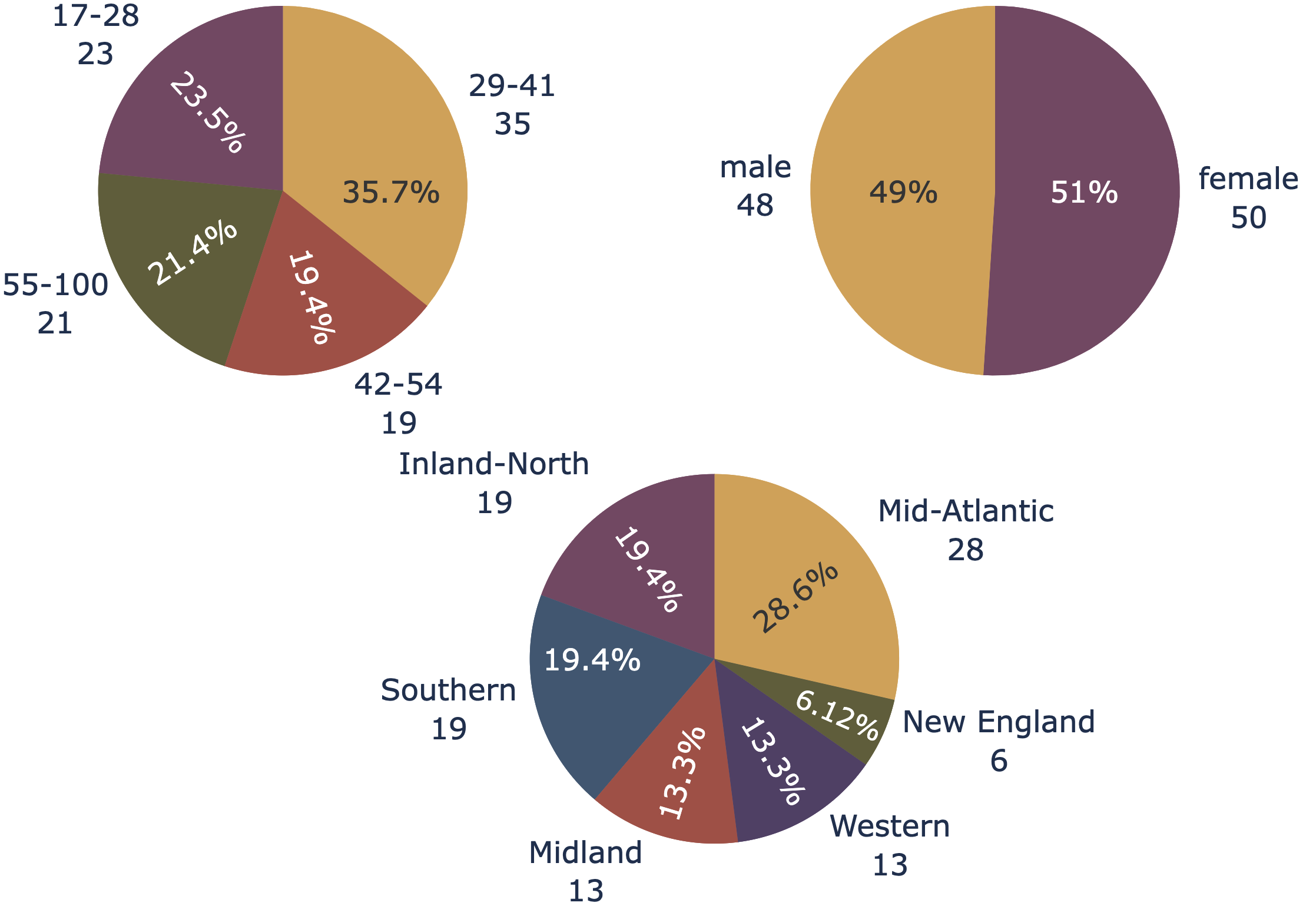}
\caption{Speaker distribution in ethnicity dataset in terms of age, dialectal region and gender.}
\label{fig:speakers-distrib-ethnicity}
\end{center}
\end{figure}

Fig.~\ref{fig:speakers-distrib-ethnicity} displays the speaker distribution in the ethnicity subset for each demographic group. 

There is no children in this dataset as their recruitment proved difficult. The distribution among the dialectal regions is similar as the one of the original test split. It is also balanced in terms of gender ($49\%$ of male and $51\%$ of female speakers). 

\section{Appendix B: Confusion bias}
\label{sec:appendix-confusion-bias}

In this section of the Appendix, we provide the complete confusion bias analysis.

\subsection{Age}

Similarly as in Table~\ref{tab:confusion-bias}, Table~\ref{tab:confusion-bias-age-group} provides the gender and dialectal region distribution for each age group in the test split. In bold, we highlight the categories for which there are less than $10\%$ of data points.

\begin{table}[H]
\centering
\resizebox{\columnwidth}{!}{%
\begin{tabular}{ccccc}
\toprule
\textbf{Age group} &
  \textbf{Samples} &
  \textbf{Speakers} &
  \textbf{Gender distribution} &
  \textbf{Dialectal region distribution} \\ \midrule
  
9-16 &
  $15788$ &
  $164$ &
  \begin{tabular}[c]{@{}c@{}}Female: $62\%$\\ Male: $38\%$\end{tabular} &
  \begin{tabular}[c]{@{}c@{}}Asian: $7\%$\\ \textbf{Inland-North: $8\%$}\\ LatinX: $10\%$\\ Mid-Atlantic: $20\%$\\ Midland: $13\%$\\ \textbf{New England: $9\%$}\\ Southern: $16\%$\\ Western: $17\%$\end{tabular} \\ \midrule
  
17-28 &
  $25414$ &
  $129$ &
  \begin{tabular}[c]{@{}c@{}}Female: $56\%$\\ Male: $44\%$\end{tabular} &
  \begin{tabular}[c]{@{}c@{}}\textbf{Asian: $7\%$}\\ Inland-North: $12\%$\\ LatinX: $13\%$\\ Mid-Atlantic: $17\%$\\ Midland: $13\%$\\ New England: $9\%$\\ Southern: $14\%$\\ Western: $15\%$\end{tabular} \\ \midrule
  
29-41 &
  $28920$ &
  $149$ &
  \begin{tabular}[c]{@{}c@{}}Female: $60\%$\\ Male: $40\%$\end{tabular} &
  \begin{tabular}[c]{@{}c@{}}\textbf{Asian: $6\%$}\\ Inland-North: $15\%$\\ LatinX: $12\%$\\ Mid-Atlantic: $17\%$\\ Midland: $14\%$\\ \textbf{New England: $7\%$}\\ Southern: $18\%$\\ Western: $12\%$\end{tabular} \\ \midrule
  
42-54 &
  $14528$ &
  $74$ &
  \begin{tabular}[c]{@{}c@{}}Female: $60\%$\\ Male: $40\%$\end{tabular} &
  \begin{tabular}[c]{@{}c@{}}\textbf{Asian: $2\%$}\\ Inland-North: $15\%$\\ \textbf{LatinX: $5\%$}\\ Mid-Atlantic: $25\%$\\ Midland: $14\%$\\ \textbf{New England: $4\%$}\\ Southern: $16\%$\\ Western: $19\%$\end{tabular} \\ \midrule
  
55-100 &
  $9854$ &
  $56$ &
  \begin{tabular}[c]{@{}c@{}}Female: $61\%$\\ Male: $39\%$\end{tabular} &
  \begin{tabular}[c]{@{}c@{}}\textbf{Asian: $1\%$}\\ Inland-North: $18\%$\\ \textbf{LatinX: $3\%$}\\ Mid-Atlantic: $28\%$\\ Midland: $14\%$\\ \textbf{New England: $6\%$}\\ Southern: $18\%$\\ Western: $12\%$\end{tabular} \\ \bottomrule
\end{tabular}%
}
\caption{Statistical distribution of audio samples for each age group in terms of gender and dialectal region in the test split.}
\label{tab:confusion-bias-age-group}
\end{table}

\subsection{Gender}

The following Table~\ref{tab:confusion-bias-gender} provides the age and dialectal region distribution for each gender in the test split. In bold, we highlight the categories for which there are less than $10\%$ of data points.

\begin{table}[H]
\centering
\resizebox{\columnwidth}{!}{%
\begin{tabular}{ccccc}
\toprule
\textbf{Gender} &
  \textbf{Samples} &
  \textbf{Speakers} &
  \textbf{Age distribution} &
  \textbf{Dialectal region distribution} \\ \midrule
  
Female &
  $55988$ &
  $336$ &
  \begin{tabular}[c]{@{}c@{}}$9-16: 17\%$\\ $17-28: 25\%$\\ $29-41: 31\%$\\ $42-54: 15\%$\\ $55-100: 11\%$\end{tabular} &
  \begin{tabular}[c]{@{}c@{}}\textbf{Asian: $6\%$}\\ Inland-North: $12\%$\\ LatinX: $11\%$\\ Mid-Atlantic: $19\%$\\ Midland: $14\%$\\ \textbf{New England: $8\%$}\\ Southern: $15\%$\\ Western: $15\%$\end{tabular} \\ \midrule
  
Male &
  $38516$ &
  $236$ &
  \begin{tabular}[c]{@{}c@{}}$9-16: 16\%$\\ $17-28: 29\%$\\ $29-41: 30\%$\\ $42-54: 15\%$\\ $55-100: 10\%$\end{tabular} &
  \begin{tabular}[c]{@{}c@{}}\textbf{Asian: $5\%$}\\ Inland-North: $15\%$\\ \textbf{LatinX: $8\%$}\\ Mid-Atlantic: $21\%$\\ Midland: $12\%$\\ \textbf{New England: $6\%$}\\ Southern: $18\%$\\ Western: $14\%$ \end{tabular} \\
\bottomrule
\end{tabular}%
}
\caption{Statistical distribution of audio samples for each gender in terms of age and dialectal region in the test split.}
\label{tab:confusion-bias-gender}
\end{table}

\subsection{Ethnicity}

Table~\ref{tab:confusion-bias-ethnicity} provides the age, gender and dialectal region distribution for each ethnicity group in the test split. 

\begin{table}[H]
\centering
\resizebox{\columnwidth}{!}{%
\begin{tabular}{ccc}
\toprule

\textbf{Ethnicity} & African American & Caucasian \\ \midrule

\textbf{Samples} & $7443$ & $7183$ \\ \midrule

\textbf{Speakers} & $50$ & $48$ \\ \midrule

\textbf{Age distrib.} & \begin{tabular}[c]{@{}c@{}}$17-28: 32\%$\\ $29-41: 42\%$\\ $42-54: 14\%$\\ $55-100:12\%$\end{tabular} & \begin{tabular}[c]{@{}c@{}}$17-28: 15\%$\\ $29-41: 29\%$\\ $42-54: 25\%$\\ $55-100: 31\%$\end{tabular} \\ \midrule

\textbf{Gender distrib.} & \begin{tabular}[c]{@{}c@{}}Female: $54\%$\\ Male: $46\%$\end{tabular} & \begin{tabular}[c]{@{}c@{}}Female: $48\%$\\ Male: $52\%$\end{tabular} \\ \midrule

\textbf{Dialectal region distrib.} & \begin{tabular}[c]{@{}c@{}}Inland-North: $20\%$\\ Mid-Atlantic: $42\%$\\ Midland: $10\%$\\ Southern: $18\%$\\ Western: $10\%$\end{tabular} & \begin{tabular}[c]{@{}c@{}} Inland-North: $19\%$\\ \\ Mid-Atlantic: $15\%$\\ Midland: $17\%$\\ New England: $12\%$\\ Southern: $20\%$\\ Western: $17\%$\end{tabular} \\ \bottomrule

\end{tabular}%
}
\caption{Statistical distribution of audio samples for each ethnicity tag in terms of age, gender and dialectal region in the test split.}
\label{tab:confusion-bias-ethnicity}
\end{table}

\section{Appendix C: Experiments -- Exhaustive statistical analysis}
\label{sec:appendix-statistical-analysis}

We display here additional results obtained with the off-the-shelf ASR model wav2vec2.0 and JointBERT SLU model, demonstrating the capabilities of our proposed dataset and methodology to quantify demographic bias in voice
assistants. Consequently, we remind the reader that the ASR and SLU models have not been particularly optimized for bias mitigation.

\subsection{Univariate tests}
\label{subsec:appendix-uni}

While results of all univariate logistic regressions were given in Section \ref{sec:results}, we also performed additional, but not mandatory, univariate tests that can be seen as complementary to the logistic regression.

\subsubsection{Logistic regression on the ethnicity subset}

The $98$ speakers for which we have an ethnicity tag recorded $14,626$ audio samples forming the ethnicity subset of the released dataset. 

Performing univariate logistic regression for variable on this subset revealed interesting results. 

No evidence of demographic bias based on \textbf{age} and \textbf{gender} was found. However, the univariate logistic regression for \textbf{dialectal region} is statistically significant at the $5\%$-level. Looking deeper in the statistical results, we see that the p-value is close to the threshold ($0.0049$) and only the coefficient of \texttt{Mid-Atlantic} has a p-value below this threshold ($0.003$). For all the other dialects, there is no evidence of statistical bias with respect to the reference category (Inland-North). Looking at the odds-ratio, speakers belonging to the Mid-Atlantic group have 0.6 times less chance to be recognized than speakers from the Inland-North group.

\subsubsection{Chi2 contingency test}

The \textbf{chi-squared} test (~\citet{pearsonChi2}) is a statistical hypothesis test used to test for the independence of several categories within a given population. However, unlike the univariate logistic regression, one cannot infer the direction of the bias since there are no coefficient or odd ratios associated to this test. 

Applied on the released test split, the \textbf{chi-squared} tests confirm what the univariate logistic regressions uncovered. All tests are statistically significant at the $5\%$-level: \textbf{gender} (p-value $=0.01$), \textbf{age} (p-value $=\num{3e-80}$), \textbf{dialect} (p-value $ \simeq 0$), \textbf{ethnicity} (p-value $=\num{4e-5}$). 

\subsubsection{One-way ANOVA test}

Another complementary univariate test is the One-way ANOVA test (notably used in~\citetlanguageresource{meyer2020artie} on the Character Error Rate). This test is also known as the "analysis of variance". It compares the means of at least 2 independent groups to assess whether there is statistical evidence that the associated population means are significantly different.

With this test, we only found significant evidence of bias for \textbf{age} (p-value $= \num{1.36e-15}$) and \textbf{dialectal region} (p-value $= \num{3.41e-44}$).

\subsection{Multivariate tests}
\label{subsec:appendix-multi}

Regarding potential mixed effects, we highlighted the ones found in Section~\ref{sec:results}. Here we provide the exhaustive and systematic analysis of all multivariate tests. 

First, gender is not a confounding factor for age: the test is statistically significant ($T=6 > q_{1,0.05}=3.84$) but the p-values of the multivariate and the univariate tests are very close, therefore we cannot conclude that gender is a confounding factor for age. The conclusion is similar for age on gender: the test is statistically significant at the $5\%$-level ($T=388 > q_{4, 0.05}=9.49$) but the p-values and odds-ratios are very close, therefore there is no confounding factor. 

Similarly, there is no evidence that age is a confounding factor for dialectal region as even if the test is statistically significant ($T = 1559 > q_{7, 0.05} = 14.07$), the p-values and odds-ratios are close, maintaining the conclusions of the univariate test unchanged. We reach the same conclusion for dialectal region with the age variable ($T=193 > q_{4,0.05} = 9.49$). 

Gender is not a confounding factor for dialectal region since the test is not significant ($T = 0.05 < q_{1,0.05}=3.84$). The test is however significant the other way around ($T = 1748 > q_{7,0.05} = 14.07$) and the coefficient for male is no longer significant at the $5\%$-level. We conclude that dialectal region is a confounding factor for gender.

Each of the previous analysis are done on the smaller ethnicity dataset in order to evaluate potential mixed effects linked with ethnicity. 

First, age is not a confounding factor for ethnicity as the test is not significant ($T=1.7 < q_{3,0.05}=7.81$). The adjustment test of ethnicity on age is significant ($T=14 > q_{1,0.05}=3.84$) but there is no change on the conclusions about age (still no significant age coefficients). 

Similarly as above, gender is not a confounding factor for ethnicity ($T=0.14 < q_{1,0.05}=3.81$). Even the adjustment test is significant the other way around ($T=16.8 > q_{1,0.05}=3.84$), conclusions for gender are unchanged (still not significant). Therefore ethnicity is not a confounding factor for gender.

Again, the conclusion is similar for dialectal region and ethnicity. The dialect is not a confounding factor for ethnicity ($T=11 < q_{5,0.05}=11.07$). And ethnicity is not a confounding factor for the dialect ($T=11 > q_{1,0.05}=3.84$ but the conclusions of the univariate test remain unchanged).

\subsection{Word Error Rate}
\label{subsec:appendix-wer}

\begin{figure*}[!h]
    \centering
    \resizebox{0.95\textwidth}{!}{
    \renewcommand{\arraystretch}{.5}
     \begin{tabular}{cc}
        \centering
         \includegraphics[width=0.8\textwidth]{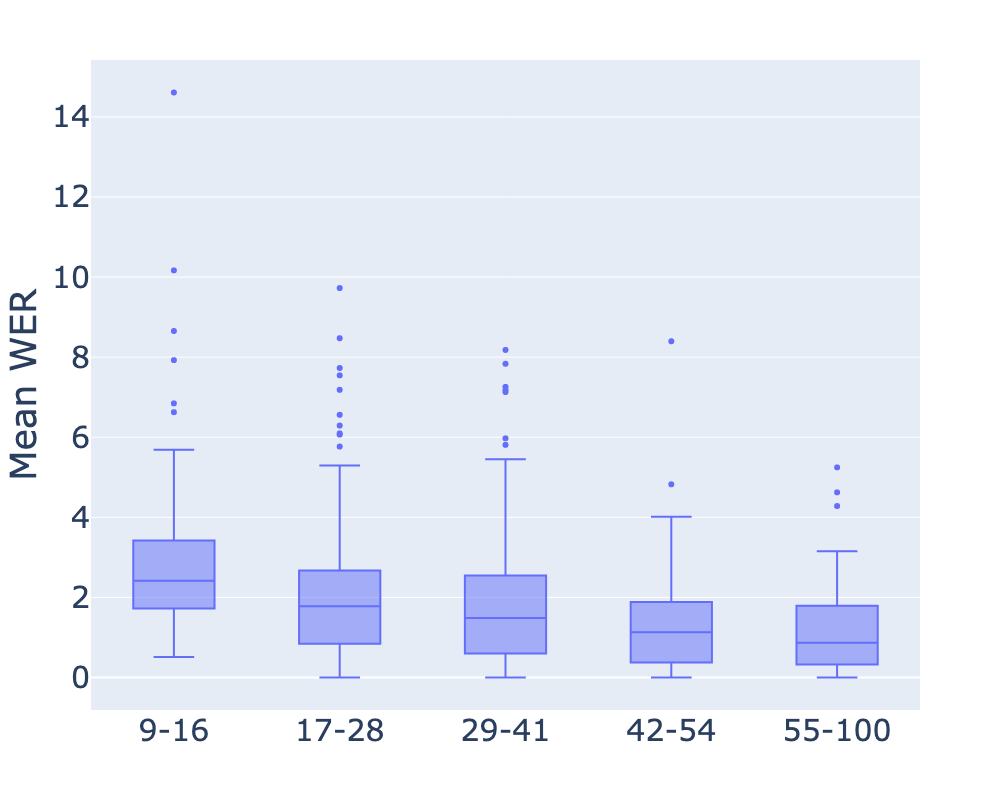} 
         &  
         \includegraphics[width=0.8\textwidth]{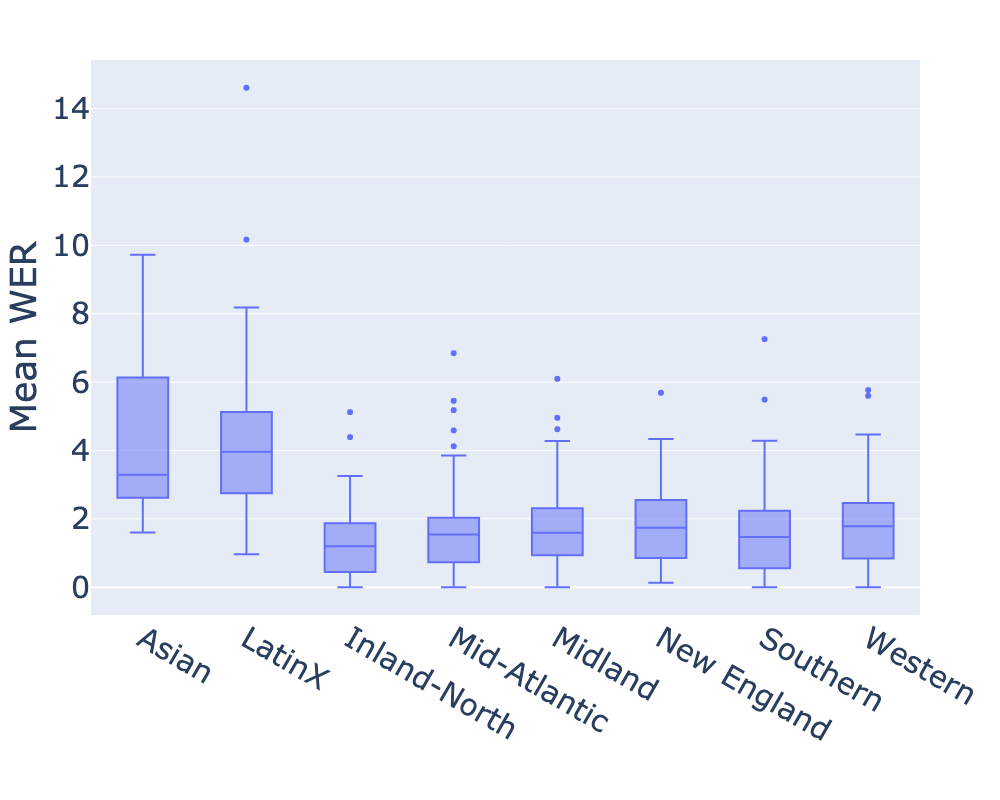} \\
         \large (a) Age group & \large (b) Dialectal region\\
         \includegraphics[width=0.8\textwidth]{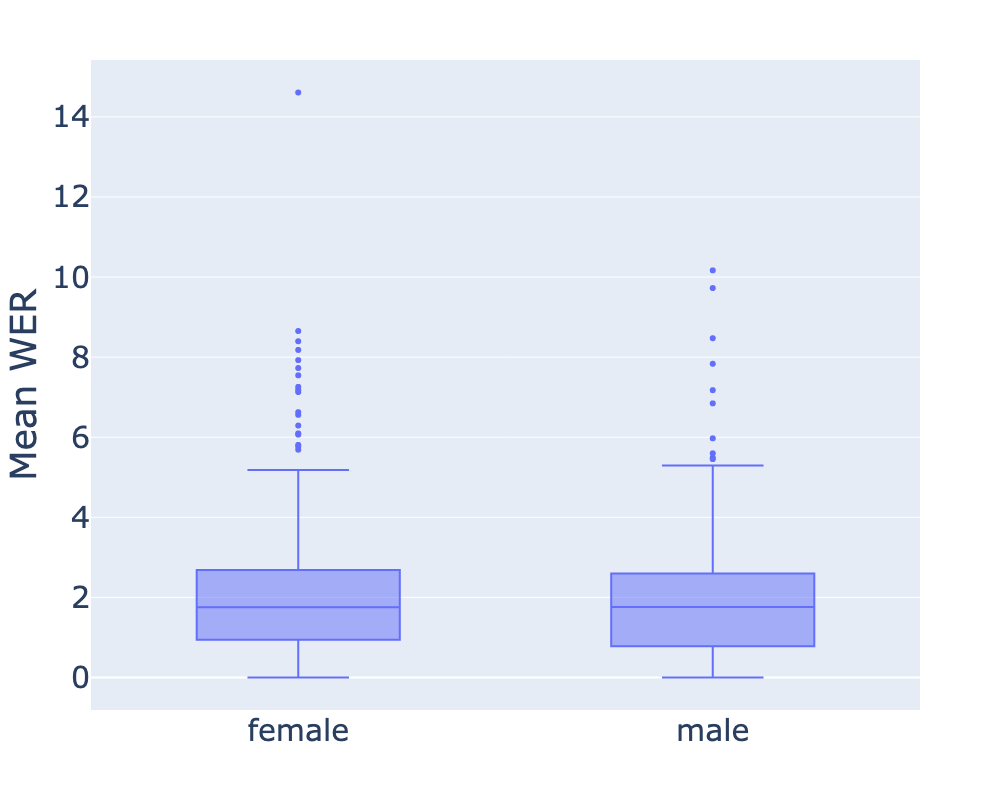} &
         \includegraphics[width=0.8\textwidth]{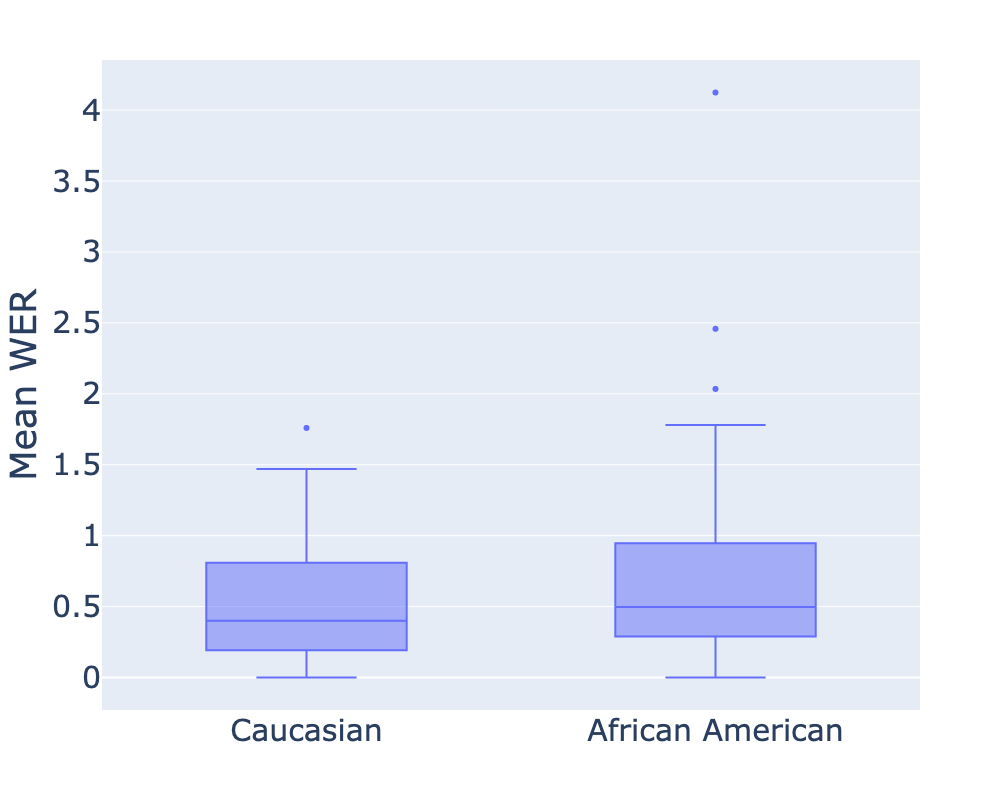} \\
         \large (c) Gender & \large (d) Ethnicity
    \end{tabular}}
    \caption{Word Error Rate (WER) per speaker's demographic group. Points indicate individual speakers.}
    \label{fig:wer-w2v}
\end{figure*}

We obtained a WER of $2.5\%$ with the fine-tuned w2v~\cite{baevski2020wav2vec}. Fig. \ref{fig:wer-w2v} displays the variation of WER per demographic group.

Similarly as in Fig. \ref{fig:emr-w2v}, there are some cases of high standard deviation.

\subsection{Speaker variability}
\label{subsec:appendix-speaker-variability}

Fig. \ref{fig:emr-w2v} in the main text, as well as Fig. \ref{fig:wer-w2v} in the previous section, showcase high standard deviations in EMR and WER per speaker. Some individual speakers (outlier points) are much less well understood than others in the same demographic group. This is consistent with the literature (for instance~\citet{tatman2017effects, tatman2020sociolinguistic} also observe much larger WERs for some individuals). ~\citet{feng2023towards} note that the recognition performance is affected by the large variability both in the pronunciation and in language use within a given speaker group.

Though it is not always possible to identify the cause just by listening to the corresponding audio clips, some speakers may have slight disfluencies or hesitations when recording. Hesitation is difficult to perfect in the world of speech recognition, as other constraints such as endpointing rules may come into play and end recognition before the user is finished speaking. Other speakers that simply utter at a slower pace may also be less consistently understood.

Lisps and other speech impediments are a third example. Most speakers in the dataset do not have any speech impediments, but it should be noted that this can greatly affect the quality of transcription. Most training data does not comprise atypical speech and therefore it is expected that these cases would be less consistently understood.

\end{document}